\documentclass{aastex631}

\usepackage{newtxtext,newtxmath,amsmath,float,xspace,threeparttable,bm,amsmath,color}

\newcommand{\todo}{\ifmmode \text{\color{red}\Huge{\(\bullet\)}} \else {\color{red}{\Huge$\bullet$}}\fi}
\newcommand{\tido}{\ifmmode {{\color{red}\bullet}} \else {\color{red}$\bullet$}\fi}

\newcommand{\E        }[1]{\ifmmode 10^{#1} \else $10^{#1}$\fi}
\newcommand{\tE        }[1]{\ifmmode \times10^{#1} \else $\times10^{#1}$\fi}
\newcommand{\til}{\ifmmode \sim \else $\sim$\fi}
\renewcommand{\~} {\ifmmode \sim \else $\sim$\fi}

\newcommand{\logNH }{\ifmmode \log (N_{\rm H}/{\rm cm}^{-2}) \else $\log (N_{\rm H}/{\rm cm}^{-2})$\fi}

\newcommand{\Mbh   }{\ifmmode M_{\rm BH} \else $M_{\rm BH}$\fi}

\newcommand{\pc}	{\ifmmode {\rm pc} \else pc\fi}
\newcommand{\ld}	{\ifmmode {\rm l.d.} \else l.d.\fi}
\newcommand{\kms}	{\ifmmode {\rm km\,s}^{-1} \else km\,s$^{-1}$\fi}
\newcommand{\cc}	{\ifmmode {\rm cm}^{-3}    \else cm$^{-3}$\fi}
\newcommand{\cmii}	{\ifmmode {\rm cm}^{-2}    \else cm$^{-2}$\fi}
\newcommand{\ergs}	{\ifmmode {\rm erg\,s}^{-1} \else erg s$^{-1}$\fi}
\newcommand{\ergcms}	{\ifmmode {\rm erg\,cm}^{-2}\,{\rm s}^{-1} \else erg\,cm$^{-2}$\,s$^{-1}$\fi}
\newcommand{\ergcmsA}	{\ifmmode {\rm erg\,cm}^{-2}\,{\rm s}^{-1}\,{\rm\AA}^{-1}
\else erg\,cm$^{-2}$\,s$^{-1}$\,\AA$^{-1}$\fi}
\newcommand{  \ergcmsHz  }{\ifmmode{\rm erg\,cm}^{-2}\,{\rm s}^{-1}\,{\rm Hz}^{-1}
                       \else ergs\,cm$^{-2}$\,s$^{-1}$\,Hz$^{-1}$\fi}
\newcommand{\kev}	{\ifmmode {\rm keV} \else keV\fi}

\newcommand{\mic}	{\ifmmode {\rm \mu m} \else $\mu$m\fi}
\newcommand{\vFWHM}	{\ifmmode v_{\mbox{\tiny FWHM}} \else $v_{\mbox{\tiny FWHM}}$\fi}
\newcommand{\vBLR}	{\ifmmode v_{\mbox{\tiny BLR}} \else $v_{\mbox{\tiny BLR}}$\fi}
\newcommand{\sigBLR}	{\ifmmode \sigma_{\mbox{\tiny BLR}} \else $\sigma_{\mbox{\tiny BLR}}$\fi}
\newcommand{\vNLR}	{\ifmmode v_{\mbox{\tiny NLR}} \else $v_{\mbox{\tiny NLR}}$\fi}
\newcommand{\tauBLR}	{\ifmmode \tau_{\mbox{\tiny BLR}} \else $\tau_{\mbox{\tiny BLR}}$\fi}

\newcommand{\Hubble}	{\ifmmode {\rm km\,s}^{-1}\,{\rm Mpc}^{-1} \else km\,s$^{-1}$\,Mpc$^{-1}$\fi}
\newcommand{\NDunit}	{\ifmmode {\rm Mpc}^{-3} \else Mpc$^{-3}$\fi}
\newcommand{\LFunit}	{\ifmmode {\rm Mpc}^{-3}\,{\rm mag}^{-1} \else Mpc$^{-3}$\,mag$^{-1}$\fi}
\newcommand{\MFunit}	{\ifmmode {\rm Mpc}^{-3}\,{\rm dex}^{-1} \else Mpc$^{-3}$\,dex$^{-1}$\fi}

\newcommand{\Msun}{\ifmmode M_{\odot} \else $M_{\odot}$\fi}
\newcommand{\Lsun}{\ifmmode L_{\odot} \else $L_{\odot}$\fi}
\newcommand{\Zsun}{\ifmmode Z_{\odot} \else $Z_{\odot}$\fi}
\newcommand{\mpyr}{\ifmmode \Msun\,{\rm yr}^{-1} \else $\Msun\,{\rm yr}^{-1}$\fi}

\newcommand{\qnote}{\ifmmode q_{0} \else $q_{0}$\fi}
\newcommand{\Hnote}{\ifmmode H_{0} \else $H_{0}$\fi}
\newcommand{\hnote}{\ifmmode h_{0} \else $h_{0}$\fi}
\newcommand{\anote}{\ifmmode a_{0} \else $a_{0}$\fi}

\newcommand{  \Halpha   }{\ifmmode {\rm H}\alpha \else H$\alpha$\fi}
\newcommand{  \ha   	}{\ifmmode {\rm H}\alpha \else H$\alpha$\fi}
\newcommand{  \Hbeta    }{\ifmmode {\rm H}\beta \else H$\beta$\fi}
\newcommand{  \hb    	}{\ifmmode {\rm H}\beta \else H$\beta$\fi}
\newcommand{  \Hgamma   }{\ifmmode {\rm H}\gamma \else H$\gamma$\fi}
\newcommand{  \Hdelta   }{\ifmmode {\rm H}\delta \else H$\delta$\fi}
\newcommand{  \Lya      }{\ifmmode {\rm Ly}\alpha \else Ly$\alpha$\fi}
\newcommand{  \Lyb      }{\ifmmode {\rm Ly}\beta \else Ly$\beta$\fi}
\newcommand{  \Pa       }{\ifmmode {\rm P}\alpha \else P$\alpha$\fi}
\newcommand{  \Pb       }{\ifmmode {\rm P}\beta \else P$\beta$\fi}
\newcommand{  \Bra      }{\ifmmode {\rm Br}\alpha \else Br$\alpha$\fi}
\newcommand{  \Brg      }{\ifmmode {\rm Br}\gamma \else Br$\gamma$\fi}
\newcommand{  \hii      }{\ifmmode {\rm H}\,\textsc{ii} \else H\,\textsc{ii}\fi}
\newcommand{  \hei      }{\ifmmode {\rm He}\,\textsc{i} \else He\,\textsc{i}\fi}
\newcommand{  \heii     }{\ifmmode {\rm He}\,\textsc{ii} \else He\,\textsc{ii}\fi}
\newcommand{  \HeIIuv   }{\ifmmode {\rm He}\,\textsc{ii}\,\lambda1640 \else He\,\textsc{ii}\,$\lambda1640$\fi}
\newcommand{  \HeIIop   }{\ifmmode {\rm He}\,\textsc{ii}\,\lambda4686 \else He\,\textsc{ii}\,$\lambda4686$\fi}
\newcommand{  \cii      }{\ifmmode {\rm C}\,\textsc{ii}  \else C\,\textsc{ii}\fi}
\newcommand{  \ciii     }{\ifmmode {\rm C}\,\textsc{iii}\right] \else C\,\textsc{iii}]\fi}
\newcommand{  \CIII     }{\ifmmode {\rm C}\,\textsc{iii}\right]\,\lambda1909 \else C\,\textsc{iii}]\,$\lambda1909$\fi}
\newcommand{  \civ      }{\ifmmode {\rm C}\,\textsc{iv}  \else C\,\textsc{iv}\fi}
\newcommand{  \CIV      }{\ifmmode {\rm C}\,\textsc{iv}\,\lambda1549 \else C\,\textsc{iv}\,$\lambda1549$\fi}
\newcommand{  \nii      }{\ifmmode [{\rm N}\,\textsc{ii}]  \else [N\,\textsc{ii}]\fi}
\newcommand{  \niii     }{\ifmmode {\rm N}\,\textsc{iii} \else N\,\textsc{iii}\fi}
\newcommand{  \niv      }{\ifmmode {\rm N}\,\textsc{iv}  \else N\,\textsc{iv}\fi}
\newcommand{  \NIVuv    }{\ifmmode {\rm N}\,\textsc{iv}\,\lambda1486 \else N\,\textsc{iv}\,$\lambda1486$\fi}
\newcommand{  \nv       }{\ifmmode {\rm N}\,\textsc{v}   \else N\,\textsc{v}\fi}
\newcommand{\oi}{\ifmmode \left[{\rm O}\,\textsc{i}\right] \else [O\,{\sc i}]\fi}
\newcommand{\OI}{\ifmmode \left[{\rm O}\,\textsc{i}\right]\,\lambda6300 \else [O\,{\sc i}]$\,\lambda6300$\fi}

\newcommand{\oii}{\ifmmode \left[{\rm O}\,\textsc{ii}\right] \else [O\,{\sc ii}]\fi}
\newcommand{\OII}{\ifmmode \left[{\rm O}\,\textsc{ii}\right]\,\lambda3727 \else [O\,{\sc ii}]\,$\lambda3727$\fi}
\newcommand{\oiii}{\ifmmode \left[{\rm O}\,\textsc{iii}\right] \else [O\,{\sc iii}]\fi}
\newcommand{\OIII}{\ifmmode \left[{\rm O}\,\textsc{iii}\right]\,\lambda5007 \else [O\,{\sc iii}]\,$\lambda5007$\fi}

\newcommand{\NII}{\ifmmode \left[{\rm N}\,\textsc{ii}\right]\,\lambda6583 \else [N\,{\sc ii}]$\,\lambda6583$\fi}

\newcommand{\NeIII}{\ifmmode \left[{\rm Ne}\,\textsc{iii}\right]\,\lambda3968 \else [Ne\,{\sc iii}]$\,\lambda3968$\fi}

\newcommand{\NeV}{\ifmmode \left[{\rm Ne}\,\textsc{v}\right]\,\lambda3426 \else [Ne\,{\sc v}]$\,\lambda3426$\fi}

\newcommand{\HeII}{\ifmmode {\rm He}\,\textsc{ii}\,\lambda4686 \else He\,{\sc ii}$\,\lambda4686$\fi}

\newcommand{\sii}{\ifmmode \left[{\rm S}\,\textsc{ii}\right] \else [S\,{\sc ii}]\fi}

\newcommand{\SII}{\ifmmode \left[{\rm S}\,\textsc{ii}\right]\,\lambda6717,6731 \else [S\,{\sc ii}]$\,\lambda6717,6731$\fi}

\newcommand{  \OIIIuv   }{\ifmmode {\rm O}\,\textsc{iii}\,\lambda1663 \else O\,\textsc{iii}\,$\lambda1663$\fi}
\newcommand{  \oiv      }{\ifmmode {\rm O}\,\textsc{iv}  \else O\,\textsc{iv}\fi}
\newcommand{  \OIVuv    }{\ifmmode {\rm O}\,\textsc{iv}\,\lambda1402  \else O\,\textsc{iv}\,$\lambda1402$\fi}
\newcommand{  \OIVIR    }{\ifmmode {\rm O}\,\textsc{iv}\,25.9\,\mu {\rm m} \else O\,\textsc{iv}\,$25.9\,\mu$m\fi}
\newcommand{  \ovi      }{\ifmmode {\rm O}\,\textsc{vi}   \else O\,\textsc{vi}\fi}
\newcommand{  \Ovi      }{\ifmmode {\rm O}\,\textsc{vi}\,\lambda1035 \else O\,\textsc{vi}\,$\lambda1035$\fi}
\newcommand{  \nei      }{\ifmmode {\rm Ne}\,\textsc{i}   \else Ne\,\textsc{i}\fi}
\newcommand{  \neii     }{\ifmmode {\rm Ne}\,\textsc{ii}  \else Ne\,\textsc{ii}\fi}
\newcommand{  \NeiiIR   }{\ifmmode {\rm Ne}\,\textsc{ii}\,12.8\,\mu {\rm m} \else Ne\,\textsc{ii}\,$12.8\,\mu$m\fi}
\newcommand{  \neiii    }{\ifmmode {\rm Ne}\,\textsc{iii} \else Ne\,\textsc{iii}\fi}
\newcommand{  \neiv     }{\ifmmode {\rm Ne}\,\textsc{iv}  \else Ne\,\textsc{iv}\fi}
\newcommand{  \nev      }{\ifmmode {\rm Ne}\,\textsc{v}   \else Ne\,\textsc{v}\fi}
\newcommand{  \NevIR    }{\ifmmode {\rm Ne}\,\textsc{v}\,24.3\,\mu {\rm m} \else Ne\,\textsc{v}\,$24.3\,\mu$m\fi}
\newcommand{  \nevi     }{\ifmmode {\rm Ne}\,\textsc{vi}  \else Ne\,\textsc{vi}\fi}
\newcommand{  \mgi      }{\ifmmode {\rm Mg}\,\textsc{i}   \else Mg\,\textsc{i}\fi}
\newcommand{  \mgii     }{\ifmmode {\rm Mg}\,\textsc{ii}  \else Mg\,\textsc{ii}\fi}
\newcommand{  \MgII     }{\ifmmode {\rm Mg}\,\textsc{ii}\,\lambda2798 \else Mg\,\textsc{ii}\,$\lambda2798$\fi}

\newcommand{  \siii     }{\ifmmode {\rm S}\,\textsc{iii} \else S\,\textsc{iii}\fi}
\newcommand{  \siv      }{\ifmmode {\rm S}\,\textsc{iv}  \else S\,\textsc{iv}\fi}
\newcommand{  \sili     }{\ifmmode {\rm Si}\,\textsc{i}   \else Si\,\textsc{i}\fi}
\newcommand{  \silii    }{\ifmmode {\rm Si}\,\textsc{ii}  \else Si\,\textsc{ii}\fi}
\newcommand{  \Siliv    }{\ifmmode {\rm Si}\,\textsc{iv}  \else Si\,\textsc{iv}\fi}
\newcommand{  \SilIVuv  }{\ifmmode {\rm Si}\,\textsc{iv}\,\lambda1400  \else Si\,\textsc{iv}\,$\lambda1400$\fi}

\newcommand{  \caii     }{\ifmmode {\rm Ca}\,\textsc{ii}   \else Ca\,\textsc{ii}\fi}
 \newcommand{\Mgb}{\ifmmode \left{\rm Mg}\,\textsc{i}\right\,\lambda5175 \else Mg\,{\sc i}\,$\lambda5175$\fi}
\newcommand{\Cahk}{\ifmmode \left[{\rm Ca H+K}\,\textsc{ii}\right\,\lambda3935,3968 \else Ca H+K$\,\lambda3935,3968$\fi}
\newcommand{  \feii     }{\ifmmode {\rm Fe}\,\textsc{ii}  \else Fe\,\textsc{ii}\fi}
\newcommand{  \feiii    }{\ifmmode {\rm Fe}\,\textsc{iii} \else Fe\,\textsc{iii}\fi}

\newcommand{ \Lhb   }{\ifmmode L\left(\hb\right) \else $L\left(\hb\right)$\fi}
\newcommand{ \fwhb  }{\ifmmode {\rm FWHM}\left(\hb\right) \else FWHM(\hb)\fi}
\newcommand{ \Lha   }{\ifmmode L\left(\ha\right) \else $L\left(\ha\right)$\fi}
\newcommand{ \fwha  }{\ifmmode {\rm FWHM}\left(\ha\right) \else FWHM(\ha)\fi}
\newcommand{ \Lmg   }{\ifmmode L\left(\mgii\right) \else $L\left(\mgii\right)$\fi}
\newcommand{ \fwmg  }{\ifmmode {\rm FWHM}\left(\mgii\right) \else FWHM(\mgii)\fi}
\newcommand{ \Lciv  }{\ifmmode L\left(\civ\right) \else $L\left(\civ\right)$\fi}
\newcommand{ \fwciv }{\ifmmode {\rm FWHM}\left(\civ\right) \else FWHM(\civ)\fi}
\newcommand{ \fwhm  }{\ifmmode {\rm FWHM} \else FWHM\fi} 
\newcommand{ \voff  }{\ifmmode v_{\rm off} \else $v_{\rm off}$\fi} 

\newcommand{ \mumg  }{\ifmmode \mu\left(\mgii\right) \else $\mu\left(\mgii\right)$\fi}
\newcommand{ \fmg   }{\ifmmode f\left(\mgii\right) \else $f\left(\mgii\right)$\fi}
\newcommand{ \muciv }{\ifmmode \mu\left(\civ\right) \else $\mu\left(\civ\right)$\fi}
\newcommand{ \fciv  }{\ifmmode f\left(\civ\right) \else $f\left(\civ\right)$\fi}

\newcommand{  \auvo     }{\ifmmode \alpha_{\nu,{\rm UVO}} \else $\alpha_{\nu,{\rm UVO}}$\fi}
\newcommand{  \Ledd     }{\ifmmode L_{\rm Edd} \else $L_{\rm Edd}$\fi}
\newcommand{  \lamLlam  }{\ifmmode \lambda L_{\lambda} \else $\lambda L_{\lambda}$\fi}
\newcommand{  \lLl      }{\ifmmode \lambda L_{\lambda} \else $\lambda L_{\lambda}$\fi}
\newcommand{  \nuLnu    }{\ifmmode \nu L_{\nu} \else $\nu L_{\nu}$\fi}
\newcommand{  \nLn      }{\ifmmode \nu L_{\nu} \else $\nu L_{\nu}$\fi}
\newcommand{  \Luv      }{\ifmmode L_{1450} \else $L_{1450}$\fi}
\newcommand{  \Lop      }{\ifmmode L_{5100} \else $L_{5100}$\fi}
\newcommand{  \lLop     }{\ifmmode \log\left(\Lop/\ergs\right) \else $\log\left(\Lop/\ergs\right)$\fi}
\newcommand{  \Lthree   }{\ifmmode L_{3000} \else $L_{3000}$\fi}
\newcommand{  \lLthree  }{\ifmmode \log\left(\Lthree/\ergs\right) \else $\log\left(\Lthree/\ergs\right)$\fi}

\newcommand{\Fthree}{\ifmmode F_{3000} \else $F_{3000}$\fi}
\newcommand{\fuv}{\ifmmode f_{\lambda}\left(1450{\rm \AA}\right) \else $f_{\lambda}\left(1450 {\rm \AA}\right)$\fi}
\newcommand{\fthree}{\ifmmode f_{\lambda}\left(3000{\rm \AA}\right) \else $f_{\lambda}\left(3000{\rm \AA}\right)$\fi}
\newcommand{\fH}{\ifmmode f_{\lambda}\left(1.65\micron\right) \else
$f_{\lambda}\left(1.65\micron\right)$\fi}

\newcommand{\fbol}{\ifmmode f_{\rm bol} \else $f_{\rm bol}$\fi}
\newcommand{\fbolwv}{\ifmmode f_{\rm bol}\left(\lambda\right) \else $f_{\rm bol}\left(\lambda\right)$\fi}
\newcommand{\fbolopt}{\ifmmode f_{\rm bol}\left(5100{\rm \AA}\right) \else $f_{\rm bol}\left(5100{\rm \AA}\right)$\fi}
\newcommand{\fbolthree}{\ifmmode f_{\rm bol}\left(3000{\rm \AA}\right) \else $f_{\rm bol}\left(3000{\rm \AA}\right)$\fi}
\newcommand{\fboluv}{\ifmmode f_{\rm bol}\left(1450{\rm \AA}\right) \else $f_{\rm bol}\left(1450{\rm \AA}\right)$\fi}

\newcommand{  \mbh      }{\ifmmode M_{\rm BH} \else $M_{\rm BH}$\fi}
\newcommand{  \lmbh     }{\ifmmode \log\left(\mbh/\Msun\right) \else $\log\left(\mbh/\Msun\right)$\fi} 
\newcommand{  \lledd    }{\ifmmode L_{\rm bol}/L_{\rm Edd} \else $L_{\rm bol}/L_{\rm Edd}$\fi}
\newcommand{  \Lbol     }{\ifmmode L_{\rm bol} \else $L_{\rm bol}$\fi}
\newcommand{  \lbol     }{\ifmmode L_{\rm bol} \else $L_{\rm bol}$\fi}
\newcommand{  \lLbol    }{\ifmmode \log\left(\Lbol/\ergs\right) \else $\log\left(\Lbol/\ergs\right)$\fi} 
\newcommand{  \Lagn     }{\ifmmode L_{\rm AGN} \else $L_{\rm AGN}$\fi}
\newcommand{  \lagn     }{\ifmmode L_{\rm AGN} \else $L_{\rm AGN}$\fi}

\newcommand{  \tgrow     }{\ifmmode t_{\rm growth} \else $t_{\rm growth}$\fi}
\newcommand{  \tUni      }{\ifmmode t_{\rm Universe} \else $t_{\rm Universe}$\fi}

\newcommand{  \Mindot	}{\ifmmode \dot{M}_{\rm infall} \else $\dot{M}_{\rm infall}$\fi}
\newcommand{  \Mbhdot	}{\ifmmode \dot{M}_{\rm BH} \else $\dot{M}_{\rm BH}$\fi}
\newcommand{  \Maddot	}{\ifmmode \dot{M}_{\rm AD} \else $\dot{M}_{\rm AD}$\fi}

\newcommand{  \as	}{\ifmmode a_{\rm *} 		\else $a_{\rm *}$\fi}
\newcommand{  \avec	}{\ifmmode \vec{a}_{\rm *} 	\else $\vec{a}_{\rm *}$\fi}
\newcommand{  \re	}{\ifmmode \eta      	\else $\eta$\fi}

\newcommand{  \mseed    }{\ifmmode M_{\rm seed} \else $M_{\rm seed}$\fi}
\newcommand{  \mbul     }{\ifmmode M_{\rm Bulge} \else $M_{\rm Bulge}$\fi} 
\newcommand{  \mstar    }{\ifmmode M_{*} \else $M_{*}$\fi} 
\newcommand{  \mgal     }{\ifmmode M_{*} \else $M_{*}$\fi} 
\newcommand{  \mhost    }{\ifmmode M_{\rm Host} \else $M_{\rm Host}$\fi}
\newcommand{  \mm       }{\ifmmode M_{*}/M_{\rm BH} \else $M_{*}/M_{\rm BH}$\fi}
\newcommand{  \mmsmall  }{\ifmmode M_{\rm BH}/M_{*} \else $M_{\rm BH}/M_{*}$\fi}
\newcommand{  \mmlarge  }{\ifmmode M_{*}/M_{\rm BH} \else $M_{*}/M_{\rm BH}$\fi}
\newcommand{  \mmwp     }{\ifmmode \left(M_{*}/M_{\rm BH}\right) \else $\left(M_{*}/M_{\rm BH}\right)$\fi}
\newcommand{  \ml       }{\ifmmode M_{*}/L_{*} \else $M_{*}/L_{*}$\fi}
\newcommand{  \mlwp     }{\ifmmode \left(M_{*}/L\right) \else $\left(M_{*}/L\right)$\fi}
\newcommand{  \mlk      }{\ifmmode \left(M_{*}/L_{K}\right) \else $\left(M_{*}/L_{K}\right)$\fi}
\newcommand{  \sigs     }{\ifmmode \sigma_{*} \else $\sigma_{*}$\fi}
\newcommand{  \Reff     }{\ifmmode R_{\rm e} \else $R_{\rm e}$\fi}

\def\kmps{\hbox{$\km\s^{-1}\,$}}

\newcommand{\bj}{\ifmmode b_{\rm J} \else $b_{\rm J}$\fi}

\newcommand{\iab}{\ifmmode i_{\rm AB} \else $i_{\rm AB}$\fi}

\newcommand{\jab}{\ifmmode J_{\rm AB} \else $J_{\rm AB}$\fi}
\newcommand{\hab}{\ifmmode H_{\rm AB} \else $H_{\rm AB}$\fi}
\newcommand{\kab}{\ifmmode K_{\rm AB} \else $K_{\rm AB}$\fi}

\newcommand{\jveg}{\ifmmode J_{\rm Vega} \else $J_{\rm Vega}$\fi}
\newcommand{\hveg}{\ifmmode H_{\rm Vega} \else $H_{\rm Vega}$\fi}
\newcommand{\kveg}{\ifmmode K_{\rm Vega} \else $K_{\rm Vega}$\fi}

\def\arcmin{\hbox{$^\prime$}}
\def\arcsec{\hbox{$^{\prime\prime}$}}

\newcommand{  \Chisq    }{\ifmmode \chi^{2} \else $\chi^{2}$}
\newcommand{  \nelec    }{\ifmmode n_{e} \else $n_{e}$\fi}     
\newcommand{  \nh       }{\ifmmode n_{\rm H} \else $n_{\rm H}$\fi}     
\newcommand{  \Ncol     }{\ifmmode N_{col} \else $N_{col}$\fi} 
\newcommand{  \NH       }{\ifmmode N_{\rm H} \else $N_{\rm H}$\fi}     

\def\sun{\hbox{$\odot$}}
\def\arcmin{\hbox{$^\prime$}}
\def\arcsec{\hbox{$^{\prime\prime}$}}

\def\ion#1#2{#1$\;${\small\rm\@Roman{#2}}\relax}

\newcommand{\OIIIa}{\ifmmode \left[{\rm O}\,\textsc{iii}\right]\,\lambda4959 \else [O\,{\sc iii}]\,$\lambda4959$\fi}
\newcommand{\NIIa}{\ifmmode \left[{\rm N}\,\textsc{ii}\right]\,\lambda6548 \else [N\,{\sc ii}]\,$\lambda6548$\fi}
\newcommand{\SIIa}{\ifmmode \left[{\rm S}\,\textsc{ii}\right]\,\lambda6716 \else [S\,{\sc ii}]\,$\lambda6716$\fi}
\newcommand{\SIIb}{\ifmmode \left[{\rm S}\,\textsc{ii}\right]\,\lambda6732 \else [S\,{\sc ii}]\,$\lambda6731$\fi}
\newcommand{\NeVa}{\ifmmode \left[{\rm Ne}\,\textsc{v}\right]\,\lambda3346 \else [Ne\,{\sc v}]\,$\lambda3346$\fi}
\newcommand{\NeVb}{\ifmmode \left[{\rm Ne}\,\textsc{v}\right]\,\lambda3426 \else [Ne\,{\sc v}]\,$\lambda3426$\fi}
\newcommand{\NeIIIa}{\ifmmode \left[{\rm Ne}\,\textsc{iii}\right]\,\lambda3869 \else [Ne\,{\sc iii}]\,$\lambda3869$\fi}
\newcommand{\NeIIIb}{\ifmmode \left[{\rm Ne}\,\textsc{iii}\right]\,\lambda3968 \else [Ne\,{\sc iii}]\,$\lambda3968$\fi}

\newcommand{\mgb}{\ifmmode \left{\rm Mg}\,\textsc{i}\right \else Mg\,{\sc i}\fi}

\def\arcmin{{\mbox{$^{\prime}$}}}

\def\arcsec{{\mbox{$^{\prime \prime}$}}}

\def\erg{{\rm\thinspace erg}}

\def\g{{\rm\thinspace g}}

\def\K{{\rm\thinspace K}}

\def\km{{\rm\thinspace km}}

\def\Lsun{\hbox{$\rm\thinspace L_{\odot}$}}
\def\m{{\rm\thinspace m}}

\def\pc{{\rm\thinspace pc}}

\def\s{{\rm\thinspace s}}

\newcommand{\halpha}{\Halpha}

\newcommand{\hbeta}{\Hbeta}

\newcommand{\HeIIir}{\ifmmode {\rm He}\,\textsc{ii}\,\lambda8237 \else He\,{\sc ii}$\,\lambda8237$\fi}
\newcommand{\HeIir}{\ifmmode {\rm He}\,\textsc{i}\,\lambda10830 \else He\,{\sc i}$\,\lambda10830$\fi}

\newcommand{\SIII}{\ifmmode \left[{\rm S}\,\textsc{iii}\right]\,\lambda9531 \else [S\,\textsc{ii}]\,$\lambda9531$\fi}

\newcommand {\Lsoftobs} {$L^{\rm obs}_{\mathrm{2-10\ keV}}$\xspace}
\newcommand {\Lbat} {$L^{\rm obs}_{\mathrm{14-195\ keV}}$\xspace}
\newcommand {\Lsoftint} {\ifmmode L^{\rm in}_{\mathrm{2-10\ keV}} \else $L^{\rm in}_{\mathrm{2-10\ keV}}$\fi}

\newcommand {\ergpersec} {\ifmmode {\rm erg~s}^{-1} \else erg~s$^{-1}$ \fi}

\newcommand {\nhunit} {cm$^{-2}$\xspace}

\def\micron{{\mbox{$\mu{\rm m}$}}}
\def\arcsec{{\mbox{$^{\prime \prime}$}}}
\def\arcmin{{\mbox{$^{\prime}$}}}

\def\arcsec{{\mbox{$^{\prime \prime}$}}}

\def\erg{{\rm\thinspace erg}}

\def\g{{\rm\thinspace g}}

\def\K{{\rm\thinspace K}}

\def\km{{\rm\thinspace km}}

\def\Lsun{\hbox{$\rm\thinspace L_{\odot}$}}
\def\m{{\rm\thinspace m}}

\def\pc{{\rm\thinspace pc}}

\def\s{{\rm\thinspace s}}

\def\ergps{\hbox{$\erg\s^{-1}\,$}}

\def\kmps{\hbox{$\km\s^{-1}\,$}}

\def\apjl{\hbox{ApJL}}
\def\apjs{\hbox{ApJS}}

\def\aap{\hbox{AAP}}
\def\micron{{\mbox{$\mu{\rm m}$}}}
\def\arcsec{{\mbox{$^{\prime \prime}$}}}
\def\arcmin{{\mbox{$^{\prime}$}}}

\newcommand {\molecfit}{\texttt{molecfit}}

\newcommand{\nuvr}{\ifmmode {\rm NUV}-r \else NUV-$r$\fi}
\newcommand{\mh}{\ifmmode M_{\rm H_2} \else $M_{\rm H_2}$\fi}
\newcommand{\mhi}{\ifmmode M_{\rm HI} \else $M_{\rm HI}$\fi}

\newcommand{\must}{\ifmmode \mu_{\ast} \else $\mu_{\ast}$\fi}
\newcommand{\hmol}{\ifmmode H_2 \else $H_2$\fi}
\newcommand{\rmol}{\ifmmode R_{\rm mol} \else $R_{\rm mol}$\fi}
\newcommand{\tdep}{\ifmmode t_{\rm dep}({\rm H_2}) \else $t_{\rm dep}({\rm H_2})$\fi}
\newcommand{\tdepHI}{\ifmmode t_{\rm dep}({\rm HI}) \else $t_{\rm dep}({\rm HI})$\fi}
\newcommand{\fgas}{\ifmmode f_{\rm H_2} \else $f_{\rm H_2}$\fi}
\newcommand{\fhi}{\ifmmode f_{\rm HI} \else $f_{\rm HI}$\fi}
\newcommand{\xco}{\ifmmode \alpha_{\rm CO} \else $\alpha_{\rm CO}$\fi}

\newcommand{\SiX}{\ifmmode \left[{\rm Si}\,\textsc{x}\right]\,\lambda14300 \else [Si\,{\sc x}]\,$\lambda14300$\fi}
\newcommand{\SiVI}{\ifmmode \left[{\rm Si}\,\textsc{vi}\right]\,\lambda19640 \else [Si\,{\sc vi}]\,$\lambda19640$\fi}
\newcommand{\SXI}{\ifmmode \left[{\rm S}\,\textsc{xi}\right]\,\lambda19196 \else [S\,{\sc xi}]\,$\lambda19196$\fi}
\newcommand{\SVIII}{\ifmmode \left[{\rm S}\,\textsc{viii}\right]\,\lambda9915 \else [S\,{\sc viii}]\,$\lambda9915$\fi}
\newcommand{\SIX}{\ifmmode \left[{\rm S}\,\textsc{ix}\right]\,\lambda12520 \else [S\,{\sc ix}]\,$\lambda12520$\fi}
\newcommand{\FeXIII}{\ifmmode \left[{\rm Fe}\,\textsc{xiii}\right]\,\lambda10747 \else [Fe\,{\sc xiii}]\,$\lambda10747$\fi}
\newcommand{\SiXI}{\ifmmode \left[{\rm Si}\,\textsc{xi}\right]\,\lambda19320 \else [Si\,{\sc xi}]\,$\lambda19320$\fi}

\newcommand{\Ndrnew}{22}
\newcommand{\Nspec}{1449} 
\newcommand{\Nspecnew}{1181} 
\newcommand{\Nnewz}{47}
\newcommand{\NAGN}{858}  

\newcommand{\Nbeamed}{105}
\newcommand{\Nunbeamed}{752}

\newcommand{\NPalomarAGN}{402}
\newcommand{\NXSHOOTERAGN}{211}

\accepted{May 4, 2022}

\submitjournal{ApJS}

\shorttitle{BASS DR2 Overview}
\shortauthors{Koss et al.}

\graphicspath{{./}{./figures/}}

\begin{document}

\title{BAT AGN Spectroscopic Survey XXI: The Data Release 2 Overview}

\correspondingauthor{Michael Koss}
\email{mike.koss@eurekasci.com}

\author[0000-0002-7998-9581]{Michael J. Koss}
\affiliation{Eureka Scientific, 2452 Delmer Street, Suite 100, Oakland, CA 94602-3017, USA}
\affiliation{Space Science Institute, 4750 Walnut Street, Suite 205, Boulder, CO 80301, USA}

\author[0000-0002-3683-7297]{Benny Trakhtenbrot}
\affiliation{School of Physics and Astronomy, Tel Aviv University, Tel Aviv 69978, Israel}

\author[0000-0001-5231-2645]{Claudio Ricci}
\affiliation{N\'ucleo de Astronom\'ia de la Facultad de Ingenier\'ia, Universidad Diego Portales, Av. Ej\'ercito Libertador 441, Santiago 22, Chile}
\affiliation{Kavli Institute for Astronomy and Astrophysics, Peking University, Beijing 100871, People's Republic of China}
 
 \author[0000-0002-8686-8737]{Franz E. Bauer}
\affiliation{Instituto de Astrof\'{\i}sica  and Centro de Astroingenier{\'{\i}}a, Facultad de F\'{i}sica, Pontificia Universidad Cat\'{o}lica de Chile, Casilla 306, Santiago 22, Chile}
\affiliation{Millennium Institute of Astrophysics (MAS), Nuncio Monse{\~{n}}or S{\'{o}}tero Sanz 100, Providencia, Santiago, Chile}
\affiliation{Space Science Institute, 4750 Walnut Street, Suite 205, Boulder, Colorado 80301, USA}

\author[0000-0001-7568-6412]{Ezequiel Treister}
\affiliation{Instituto de Astrof{\'i}sica, Facultad de F{\'i}sica, Pontificia Universidad Cat{\'o}lica de Chile, Casilla 306, Santiago 22, Chile}

\author[0000-0002-7962-5446]{Richard Mushotzky}
\affiliation{Department of Astronomy, University of Maryland, College Park, MD 20742, USA}
\affiliation{Joint Space-Science Institute, University of Maryland, College Park, MD 20742, USA}

\author[0000-0002-0745-9792]{C. Megan Urry}
\affiliation{Yale Center for Astronomy \& Astrophysics and Department of Physics, Yale University, P.O. Box 208120, New Haven, CT 06520-8120, USA}
 
\author[0000-0001-8211-3807]{Tonima T. Ananna}
\affiliation{Department of Physics and Astronomy, Dartmouth College, 6127 Wilder Laboratory, Hanover, NH 03755, USA}

\author[0000-0003-0476-6647]{Mislav Balokovi\'c}
\affiliation{Yale Center for Astronomy \& Astrophysics, 52 Hillhouse Avenue, New Haven, CT 06511, USA}
\affiliation{Department of Physics, Yale University, P.O. Box 2018120, New Haven, CT 06520, USA}

\author[0000-0002-8760-6157]{Jakob S. den Brok}
\affiliation{Institute for Particle Physics and Astrophysics, ETH Z{\"u}rich, Wolfgang-Pauli-Strasse 27, CH-8093 Z{\"u}rich, Switzerland}
\affiliation{Argelander Institute for Astronomy, Auf dem H{\"u}gel 71, 53231, Bonn, Germany}

\author[0000-0003-1673-970X]{S. Bradley Cenko}
\affiliation{Astrophysics Science Division, NASA Goddard Space Flight Center, Mail Code 661, Greenbelt, MD 20771, USA}
\affiliation{Joint Space-Science Institute, University of Maryland, College Park, MD 20742, USA}

\author{Fiona Harrison}
\affiliation{Cahill Center for Astronomy and Astrophysics, California Institute of Technology, Pasadena, CA 91125, USA}

\author[0000-0002-4377-903X]{Kohei Ichikawa}
\affil{Frontier Research Institute for Interdisciplinary Sciences, Tohoku University, Sendai 980-8578, Japan}
\affil{
Astronomical Institute, Tohoku University, Aramaki, Aoba-ku, Sendai, Miyagi 980-8578, Japan
}
\affil{Max-Planck-Institut f{\"u}r extraterrestrische Physik (MPE), Giessenbachstrasse 1, D-85748 Garching bei M{\"u}unchen, Germany
}

\author[0000-0003-3336-5498]{Isabella Lamperti}
\affiliation{Centro de Astrobiología (CAB), CSIC–INTA, Cra. de Ajalvir Km. 4, 28850 Torrejón de Ardoz, Madrid, Spain}

\author[0000-0002-7851-9756]{Amy Lein}
\affiliation{Astrophysics Science Division, NASA Goddard Space Flight Center, Mail Code 661, Greenbelt, MD 20771, USA}

\author[0000-0001-8450-7463]{Julian E. Mej\'ia-Restrepo}
\affiliation{European Southern Observatory, Casilla 19001, Santiago 19, Chile}

\author[0000-0002-5037-951X]{Kyuseok Oh}
\affiliation{Korea Astronomy \& Space Science institute, 776, Daedeokdae-ro, Yuseong-gu, Daejeon 34055, Republic of Korea}
\affiliation{Department of Astronomy, Kyoto University, Kitashirakawa-Oiwake-cho, Sakyo-ku, Kyoto 606-8502, Japan}
\affiliation{JSPS Fellow}

\author[0000-0001-9879-7780]{Fabio Pacucci}
\affil{Center for Astrophysics $\vert$ Harvard \& Smithsonian,
Cambridge, MA 02138, USA}
\affil{Black Hole Initiative, Harvard University,
Cambridge, MA 02138, USA}

\author[0000-0001-8640-8522]{Ryan W. Pfeifle}
\affiliation{Department of Physics and Astronomy, George Mason University, 4400 University Drive, MSN 3F3, Fairfax, VA 22030, USA}

\author[0000-0003-2284-8603]{Meredith C. Powell}
\affiliation{Kavli Institute of Particle Astrophysics and Cosmology, Stanford University, 452 Lomita Mall, Stanford, CA 94305, USA}

\author[0000-0003-3474-1125]{George C. Privon}
\affiliation{National Radio Astronomy Observatory, 520 Edgemont Road, Charlottesville, VA 22903, USA}
\affiliation{Department of Astronomy, University of Florida, P.O. Box 112055, Gainesville, FL 32611, USA}

\author[0000-0001-5742-5980]{Federica Ricci}
\affiliation{Instituto de Astrof{\'i}sica, Facultad de F{\'i}sica, Pontificia Universidad Cat{\'o}lica de Chile, Casilla 306, Santiago 22, Chile}
\affiliation{Dipartimento di Fisica e Astronomia, Università di Bologna, via Gobetti 93/2, 40129 Bologna, Italy}

\author[0000-0001-7116-9303]{Mara Salvato}
\affiliation{Max-Planck-Institut f{\"u}r extraterrestrische Physik (MPE), Giessenbachstrasse 1, D-85748 Garching bei M{\"u}unchen, Germany}

\author[0000-0001-5464-0888]{Kevin Schawinski}
\affiliation{Modulos AG, Technoparkstrasse 1, CH-8005 Zurich, Switzerland}

\author[0000-0002-2125-4670]{Taro Shimizu}
\affiliation{Max-Planck-Institut f{\"u}r extraterrestrische Physik (MPE), Giessenbachstrasse 1, D-85748 Garching bei M{\"u}unchen, Germany}

\author[0000-0001-5785-7038]{Krista L. Smith}
\affiliation{Department of Physics, Southern Methodist University, 3215 Daniel Ave., Dallas, TX 75205, USA}

\author[0000-0003-2686-9241]{Daniel Stern}
\affiliation{Jet Propulsion Laboratory, California Institute of Technology, 4800 Oak Grove Drive, MS 169-224, Pasadena, CA 91109, USA}

 
\begin{abstract}
The BAT AGN Spectroscopic Survey (BASS) is designed to provide a highly complete census of the key physical parameters of the supermassive black holes (SMBHs) that power local active galactic nuclei (AGN) ($z\lesssim0.3$), including their bolometric luminosity (\Lbol), black hole mass (\mbh), accretion rates (\lledd), line-of-sight gas obscuration (\NH), and the distinctive properties of their host galaxies (e.g., star formation rates, masses, and gas fractions).  
We present an overview of the BASS data release 2 (DR2), an unprecedented spectroscopic AGN survey in spectral range, resolution, and sensitivity, including \Nspec\ optical ($\sim$3200~\AA-1~\mic) and 233 NIR (1-2.5~\mic) spectra for the brightest \NAGN\ ultra-hard X-ray (14-195 keV) selected AGN across the entire sky and essentially all levels of obscuration. 
This release provides a highly complete set of key measurements (emission line measurements and central velocity dispersions), with 99.9\% measured redshifts and 98\% black hole masses estimated (for unbeamed AGN outside the Galactic plane).
The BASS DR2 AGN sample represents a unique census of nearby powerful AGN, spanning over 5 orders of magnitude in 
AGN bolometric luminosity ($\lbol\sim10^{40}-10^{47}\,\ergs$), 
black hole mass ($\mbh\sim10^{5}-10^{10}\,\Msun$),
Eddington ratio ($\lledd\gtrsim 10^{-5}$), 
and obscuration ($\NH\sim 10^{20}-10^{25}\,\cmii$).
The public BASS DR2 sample and measurements can thus be used to answer fundamental questions about SMBH growth and its links to host galaxy evolution and feedback in the local universe, as well as open questions concerning SMBH physics.
Here we provide a brief overview of the survey strategy, the key BASS DR2 measurements, data sets and catalogs, and scientific highlights from a series of DR2-based works pursued by the BASS team.
\end{abstract}



\section{Introduction} 
\label{sec:intro}

Although active galactic nuclei (AGNs) and the supermassive black holes (SMBHs) that power them have been studied for decades, there still many key unresolved questions concerning the nature of these systems and how their evolution may be related to the galaxies that host them.
While there are some clear SMBH-host correlations, such as the one between SMBH mass and bulge properties \citep{Kormendy:2013:511} or the similar redshift evolution of star formation and SMBH growth \cite[e.g.,][]{Heckman:2014:589,2019MNRAS.485.3721Y}, it is not yet clear whether SMBH growth and AGN output (``feedback'', e.g., radiation, winds, jets) affects the host galaxy interstellar medium, star formation, and molecular gas \citep[e.g.,][]{Fabian:2012:455a}.  On smaller scales, it is not yet entirely clear what is the structure of the obscuring torus, what is its connection to its surroundings \citep[e.g.,][]{Netzer:2015:365} and what is the role of obscuration by galaxy-scale gas and nuclear starbursts.  Moreover, the stochasticity of both SMBH fueling and corresponding AGN emission \citep[e.g.,][]{Hickox:2014:9,Schawinski:2015:2517}, combined with inherent biases in AGN survey techniques when identifying obscured AGN \citep[e.g.,][]{Hickox:2018:625}. 
Therefore, the only way to address the many open questions concerning SMBHs and AGNs, even in the local universe, and to construct statistically significant global trends and/or correlations, necessitates large samples (e.g. $N>100$) surveyed with multiwavelength observations.

Extragalactic hard (${>}2$\,keV) X-ray surveys provide one of the most complete ways to study growing black holes in an unbiased way \citep[see, e.g.,][]{Brandt:2015:1a,2021arXiv211101156B}.  A large fraction, and indeed the majority, of the AGN population is obscured and therefore the construction of a complete AGN census requires the identification of both obscured and unobscured sources \citep[see review by, e.g.,][]{Hickox:2018:625}. At even higher energies, the ultra-hard X-rays ($>10$\,keV) provide a more complete tracer of the radiation for obscured AGN ($\NH > 10^{22}\,\cmii$) and even some Compton-thick (CT) AGN \citep[$\NH > 10^{24}\,\cmii$; e.g.,][]{Ricci:2015:L13,Koss:2016:85}.  An all-sky survey in the ultra-hard X-ray band (${>}$10 keV) thus provides an important way to answer the fundamental questions of SMBH growth and its links to host galaxy evolution, as well as many open questions concerning AGN physics, for a complete, unbiased sample of AGN.  


Over the past 20 years, great progress has been made in surveying the ultra-hard X-ray sky to increasing depths with the Burst Alert Telescope \citep[BAT;][]{Barthelmy:2005:143} at 14-195 keV on board the Neil Gehrels Swift Observatory \citep{Gehrels:2004:1005} and the IBIS instrument \citep{Ubertini:2003:L131} at 17-60 keV onboard the INTEGRAL observatory \citep{Winkler:2003:L1}.  Thanks to its wide field of view (FOV; 1.4 sr half coded), BAT monitors roughly 80\% of the sky every day, providing regularly sampled average emission properties of objects.  INTEGRAL/IBIS, with a roughly 13 times smaller FOV (0.11 sr half coded) but better angular resolution (12\arcmin\ vs. 19.5\arcmin\ for BAT), has focused on targeting the Galactic plane and particularly the Galactic center region.  Thus, the Swift BAT survey provides a uniform all-sky census of the average ultra-hard X-ray emission of AGN.

Early analysis \citep[e.g.,][]{Markwardt:2005:L77,Tueller:2008:113} of the (ongoing) all-sky survey with BAT found that the brightest ultra-hard-X-ray-selected AGN in the sky can probe nearby ($z<0.1$) AGN, including highly obscured systems; low-redshift, high-luminosity AGN and quasars ($0.05 \lesssim z \lesssim 0.3$), and much more distant, beamed AGN (reaching out to $z\gtrsim3$).
Specifically, the unbeamed BAT-detected AGN span the moderate-to-high-luminosity end of the X-ray luminosity function (XLF; e.g., \citealp{Maccacaro:1991:117}, \citealp{Comastri:1995:1}, \citealp{Gilli:2007:79}, \citealp{Ueda:2014:104}, \citealp{Aird:2015:66}, \citealp{Buchner:2015:89}, \citealp{Ananna:2019:240}) at all accessible redshifts, thus providing critical low-redshift templates for high-$z$ sources detected in much deeper, pencil-beamed surveys (e.g., CDF-S, \citealp{Luo:2017:2}; COSMOS, \citealp{Civano:2016:62}).  
The low-redshift regime of the BAT AGN survey thus strongly complements high-redshift AGN surveys, providing a legacy dataset with both high spatial resolution\footnote{The angular resolution at the median redshift $z{=}0.04$ is 10${\times}$ sharper than at $z{=}1$, which is typical for deeper and narrower surveys.} and very high signal-to-noise ratio data obtained through relatively short observations.  
Finally, some AGN are uniquely identified as AGN in the hard X-rays \citep[e.g.,][]{Smith:2014:112}. Other all-sky selection methods, such as WISE mid-IR (MIR) colors, are only able to uniformly classify the most luminous AGN  (\Lsoftobs${>}10^{44}\,\ergs$; e.g., \citealp{Ichikawa:2017:74}) owing to contamination from star formation in the host galaxies.  
Moreover, methods that use strong emission-line ratio diagnostics to identify AGN \citep[e.g., ``BPT''][]{Baldwin:1981:5, Veilleux:1987:295, Kewley:2006:961} only select about half of the ultra-hard X-ray-selected BAT AGN \citep{Koss:2017:74}. This is likely due to the contribution of star forming regions in the AGN host galaxies \citep[e.g.,][]{Koss:2011:L42}, combined with dust obscuration of the central, AGN-dominated line-emitting regions \citep[e.g.,][]{Koss:2010:L125}.

The ultimate goal of the BAT AGN Spectroscopic Survey (BASS) is to provide the largest available spectroscopic sample of Swift BAT ultra-hard X-ray (14--195\,keV) detected AGN. BASS DR1 \citep{Koss:2017:74} used mostly archival optical spectra for 641 AGN from the 70-month BAT catalog \citep{Baumgartner:2013:19}.  This was complemented by detailed 0.5--200 keV spectral measurements using Swift, Chandra, and  XMM-Newton for 838 AGN \citep{Ricci:2017:17}. A further 102 near-IR (NIR; \hbox{1--2.5\,\micron}) spectra of BAT AGN were reported on by \citet{Lamperti:2017:540}.  Other data include high spatial resolution NIR adaptive optics (AO) imaging \citep{Koss:2018:214a}; extensive continuum modeling of the MIR and far-IR (FIR) emission from WISE, IRAS, Akari, and Herschel \citep{Ichikawa:2017:74, Ichikawa:2019:31}; radio emission in several resolution regimes \citep{Baek:2019:4317,Smith:2020:4216}; and host-scale molecular gas measurements \citep{Koss:2021:29}. From these data, numerous links and correlations were investigated, such as between X-ray emission and high-ionization optical lines \citep{Berney:2015:3622}, and/or ionized gas outflows \citep{Rojas:2020:5867}. A major result was the realization that the Eddington ratio is a key parameter in some of these links and scaling relations \citep[e.g.,][]{Ricci:2017:488,Oh:2017:1466,Ricci:2018:1819}.  Other BASS DR1 studies focused on AGN clustering \citep{Powell:2018:110}, the surprisingly weak (or indeed insignificant) correlation between the X-ray photon index, $\Gamma$, and Eddington ratio \citep{Trakhtenbrot:2017:800},  the most luminous obscured AGN \citep{Bar:2017:2879}, optically unobscured AGN with massive X-ray absorbing columns \citep{Shimizu:2018:154}, BAT blazars \citep{Paliya:2019:154}, and a search for BH binaries \citep{Liu:2020:122}.

Since making BASS DR1 public, the BASS team has pursued additional optical and NIR spectroscopy towards the second data release of BASS (DR2), to obtain higher spectral resolution ($R\gtrsim1500$) over the broadest possible spectral range (i.e., 3200--10000\,\AA), for the most complete sample of AGN drawn from the 70-month BAT catalog.  
Below we present an overview of the BASS DR2, including a summary of our survey design and goals, datasets, key scientific results, and a comparison to other large AGN surveys.  
A more detailed, technical description of the DR2 observations and data is provided in \citet{Koss_DR2_catalog}.

Throughout this work we adopt $\Omega_{\rm M} {=} 0.3$, $\Omega_\Lambda {=} 0.7$, and $H_0 {=} 70\,\kms\,{\rm Mpc}^{-1}$.  To determine the extinction due to Milky Way foreground dust, we use the maps of \citet{Schlegel:1998:525} and the extinction law derived by \citet{Cardelli:1989:245}.

\section{Overview of DR2 Special Issue}
\label{sec:overview_issue}

BASS DR2 provides optical spectra and associated redshifts, broad and narrow emission line measurements, velocity dispersions, and derived quantities --- particularly BH mass and Eddington ratio estimates --- for a nearly complete (e.g. $>$95\%) sample of the \NAGN\ AGN from the Swift BAT 70-month survey.  


\subsection{Major Science Goals and Key measurements}
\label{sec:science_goals}

The measurements and derived quantities provided through the BASS project are critical for the major science goals of the survey:

\begin{enumerate}
    \item {\it Provide a (nearly) complete census of the brightest local AGN from the unobscured to highly obscured}.\\  
    Swift BAT, with its ultra hard X-ray sensitivity, serves as a primary discovery and survey tool for these local AGN as the energy range is uncontaminated by star formation and (nearly) unaffected by obscuration, and hence naturally provides a reliable tracer of AGN emission and SMBH growth within a volume-limited, highly complete survey.  
    
    \item {\it Connect BH growth to their host galaxy properties to understand fueling and/or feedback}.\\  
    It is critical to understand how the large range of SMBH-related properties --- such as the bolometric luminosities, BH masses (\mbh), accretion rates (\lledd), kinetic power of AGN-powered outflows and jets, and circumnuclear obscuration --- relate to key host galaxy properties as traced by multiwavelength data, such as star formation rates (SFRs), stellar and molecular gas content, morphologies, and merger activity. 
    
    \item {\it Provide critical nearby templates for luminous high-redshift AGN}.\\  
    The BASS AGN are local analogs of the powerful AGN that form the bulk of the X-ray-detected population in deep pencil-beam surveys where high spatial resolution (e.g., hundreds of parsecs) and sensitivity cannot be achieved.  
    
    \item {\it Provide critical diagnostics for rare and/or ``abnormal'' AGN}.\\  
    The unprecedentedly rich collection of multiwavelength data collected for the BASS AGN and its unique selection in the ultra-hard X-rays allow one to identify, calibrate, and/or test selection criteria for highly obscured AGN, highly accreting AGN (i.e., $\lledd\gtrsim1$), and other challenging subclasses, to be used with future facilities, surveys, and models.
    
    
\end{enumerate}

\subsection{Revised 70-month AGN Catalog}
\label{sec:revised_catalog}

We briefly review the AGN catalog changes compared to  BASS DR1, with further details provided in \citet{Koss_DR2_catalog}.  The published 70-month BAT catalog \citep{Baumgartner:2013:19} is composed of 1210 ultra-hard X-ray sources, including 822 classified as AGN or associated with a galaxy and likely an AGN, 287 Galactic sources (e.g., high-mass X-ray binary, low-mass X-ray binary, cataclysmic variable, pulsar), 19 clusters, and 82 unknown sources.  In the BASS DR1 \citep[][see Appendix A]{Ricci:2017:17} new AGN candidates were identified among BAT detections based on WISE and soft X-ray data to increase the number of 70-month AGN to 838.

However, even after the DR1, 44 unknown BAT sources, typically near the Galactic plane ($\lvert b\rvert{<}10^{\circ}$), had not been associated with counterparts.  Of these sources, \Ndrnew, were found to be AGN.  Fifteen of these unknown sources were subsequently identified as Galactic with NuSTAR and/or Chandra follow-up surveys \citep[e.g.,][]{Yukita:2017:47,Kennedy:2020:3912,Halpern:2018:247} and/or based on optical spectroscopy obtained in the BASS/DR2 \citep[see][for details]{Koss_DR2_catalog}.  Another two sources that were classified as AGN in the DR1 based solely on their hard X-ray spectra were found to be Galactic.  

Unfortunately, their were still seven sources that lie very close to the Galactic plane ($\lvert b\rvert{<}3^{\circ}$) owing to their very high extinction values ($A_V \sim 5-43$ mag) and very high source confusion in the optical/NIR (i.e., multiple stars per  1$\arcsec$ sq area), made optical and NIR spectroscopy of the counterpart impractical. The number of AGN is \NAGN\ after these updates since two sources in the DR1 were discovered to be Galactic.\footnote{838 DR1 AGN - 2 DR1 AGN found to be Galactic + 22 New AGN= 858 DR2 AGN + 7 unknown sources at $\lvert b\rvert{<}3^{\circ}$} 

For consistency with BASS DR1 and earlier studies, we classify AGN according to the presence (or absence) of broad emission lines. 
Specifically, Sy 1 are AGN with broad \hbeta\ line emission, Sy 1.9 have narrow \hbeta\ and broad \halpha, while Sy 2 AGN have both narrow \hbeta\ and narrow \halpha\ (including small numbers of LINERs and AGN in H$_2$-dominated regions).
AGN type based on optical spectroscopy. For beamed AGN, the types include those with the presence of broad lines (BZQ), only host galaxy features lacking broad lines (BZG), or traditional continuum-dominated blazars, with no emission lines or host galaxy features (BZB).

In addition to unbeamed AGN, the Swift BAT survey includes also beamed and/or lensed AGN, which are important to separate for most scientific analyses.  
The DR2 has \Nbeamed\ beamed AGN as identified by their multiwavelength emission, and particularly radio emission, and/or Gamma-ray emission detected by Fermi \citep[e.g.,][]{Paliya:2019:154}.  This includes both blazars where the boosted continuum emission completely dominates the (rest-frame) UV/optical regime and no significant emission lines are seen, and flat-spectrum radio quasars (FSRQs), which do show broad emission lines \citep[e.g.,][]{Paliya:2019:154}.  There are additionally two lensed AGN. One of the beamed AGN (SWIFT J1833.7-2105, aka PKS 1830-211 at $z{=}2.5$) is also lensed \citep{Lidman:1999:L57}) by a foreground galaxy.  Thus, the total sample of \NAGN\ includes \Nunbeamed\ unbeamed AGN, 104 beamed AGN, 1 beamed and lensed AGN, and 1 lensed and unbeamed AGN.

\begin{figure*}[hbtp]
\centering
\includegraphics[width=\textwidth]{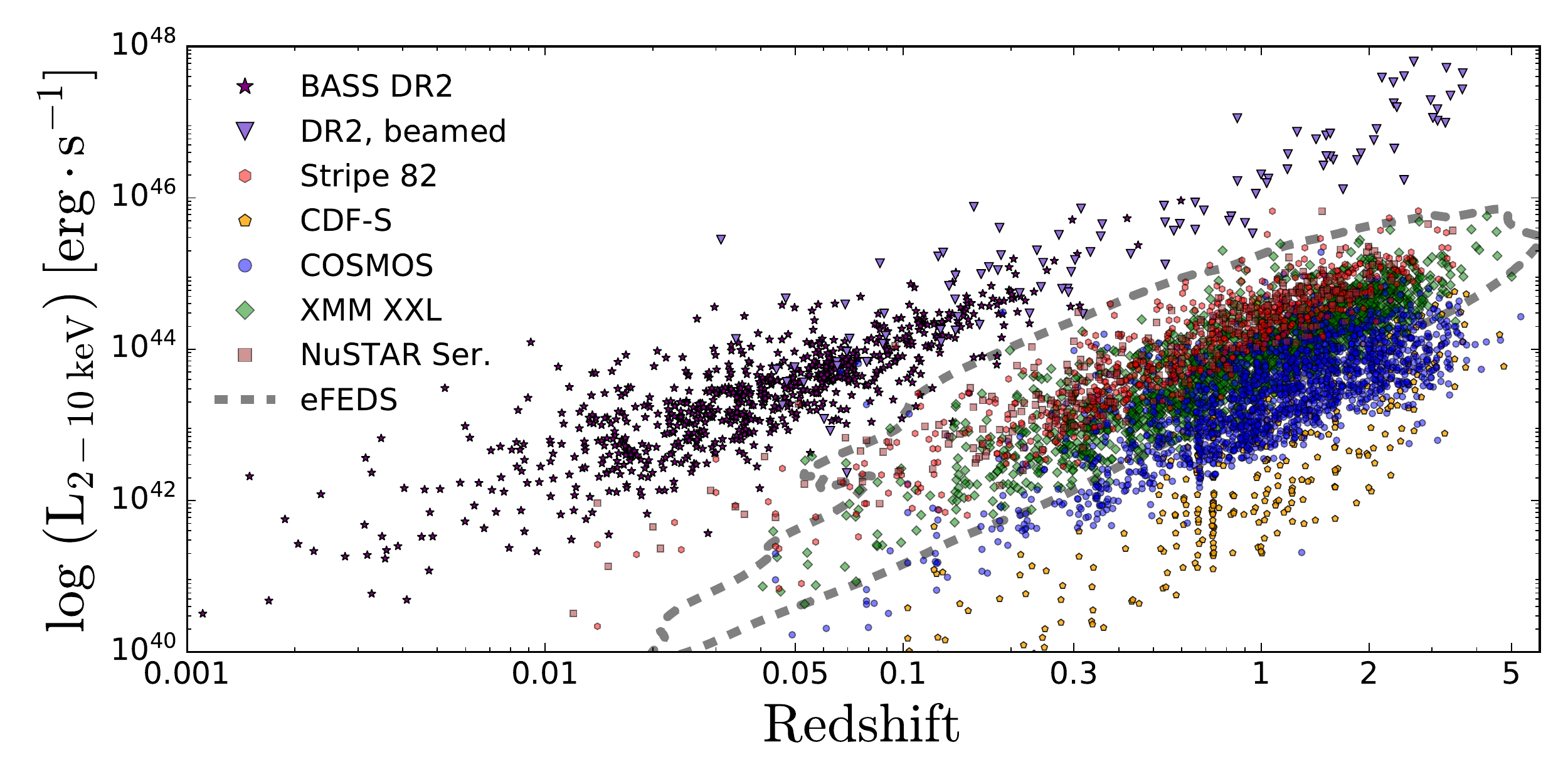}
\caption{The rest-frame 2--10\,keV luminosities of the BASS DR2 AGN and of higher-redshift X-ray AGN surveys. Although BASS AGN are selected based on their 14--195 keV emission, we plot the best-fit intrinsic 2--10 keV emission based on detailed X-ray spectral modeling  \citep{Ricci:2017:488}.  
Unbeamed BASS AGN are shown with purple stars, while beamed AGN are shown with purple triangles.    
The unbeamed AGN in the BASS sample tend to span the moderate-to-high-luminosity end of the XLF. We also show samples drawn from deeper X-ray AGN surveys, including Stripe 82X \citep[red pentagons;][]{LaMassa:2016:172,Ananna:2017:66}; CDF-S \citep[brown pentagons;][]{Luo:2017:2}; Chandra COSMOS Legacy survey \citep[blue circles;][]{Civano:2016:62, Marchesi:2016:34};
XMM XXL \citep[green diamonds;][]{Pierre:2016:A1}; and the NuSTAR Serendipitous Survey \citep[yellow squares;][]{Lansbury:2017:99}.  We also show contours for eFEDS containing 99\% of the data \citep[][dashed grey,]{2021arXiv210614520S}.  We limit our comparison to X-ray sources with confirmed counterparts with spectroscopic redshifts.    For eFEDS, unlike the other surveys, the soft-band flux (0.2-2.3 keV) was used to estimate the hard X-ray 2-10 keV emission assuming a power-law spectral model with $\Gamma=1.7$ since only a small number of sources were detected above 2.3 keV \citep[$<$1\%][]{2021arXiv210614517B}. For these higher-redshift X-ray surveys we assumed a power-law spectral model with $\Gamma=1.7$, to bring each X-ray luminosity to the rest frame by $K$-correcting the apparent luminosities based on the observed redshifts into the 2-10 keV rest frame.
The deeper, higher-redshift samples tend to sample a luminosity range that is consistent with that covered by BAT, but at higher redshift.}
\label{fig:Lum_xray}
\end{figure*}

An X-ray luminosity and redshift plot of the BASS DR2, with the newly revised redshifts and AGN classifications, is shown in \autoref{fig:Lum_xray}.  The figure also shows a few other deep, and narrow X-ray AGN surveys. We discuss these samples and comparison in more detail in \autoref{sec:other_surveys}; however, it is clearly evident that BASS provides a natural low-redshift benchmark for distant surveys.

\subsection{Survey Strategy and Observations}
\label{sec:strategy_obs}

\begin{figure*} 
\centering
\includegraphics[width=\textwidth]{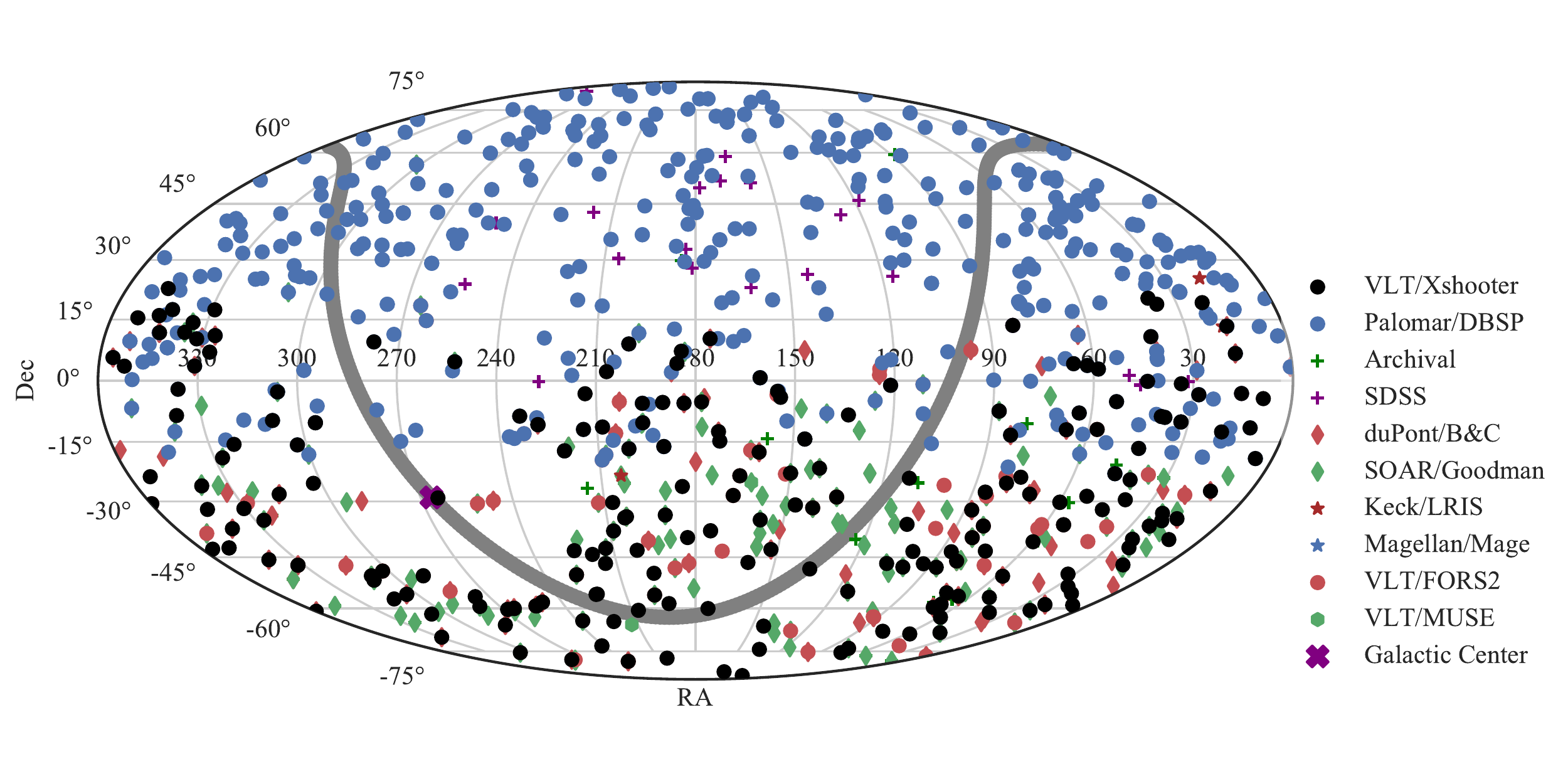}
\caption{Overview of the BASS DR2 optical spectroscopy as observed on the sky (shown in equatorial coordinates and a Mollweide projection).  The Galactic plane is indicated by the light grey line.}
\label{fig:spectra_map}
\end{figure*}

The sky distribution of all BASS DR2 optical spectroscopic observations is presented in \autoref{fig:spectra_map}.  \autoref{tab:bass_dr2} summarizes all of the BASS DR2 data, with further details provided in \citet{Koss_DR2_catalog}.  The majority of the spectra used for the catalog measurements presented in DR2 papers come from either the Double Beam Spectrograph (DBSP) mounted on the Palomar Hale 5 m telescope (\NPalomarAGN\ AGN, mainly in the northern hemisphere) or the X-Shooter spectrograph at the Very Large Telescope (VLT; \NXSHOOTERAGN\ sources, mainly southern).

In terms of DR2 spectroscopic targeting, our goals were to (1) provide the largest possible sample of black hole mass measurements from either broad Balmer emission lines or stellar velocity dispersion measurements and (2) cover broadest possible spectral range (e.g., 3200--10000\,\AA) for emission-line measurements, for the entire catalog of \NAGN\ AGN.  
In practice, unless one uses echelle instruments (e.g., X-Shooter), the latter goal requires either spectra with broad wavelength coverage and lower resolution or, alternatively, multiple, higher-resolution gratings and narrower wavelength coverage.
The former goal motivates a spectral resolution of $R\gtrsim1000$ ($\Delta v \lesssim 300\,\kms$ for broad lines), and yet higher resolution ($>$2000) for stellar velocity dispersion measurements.  Our targeting strategy was further complicated by the fact that for a significant number of targets we did not know \textit{a priori} the broad/narrow-line nature of the sources, and thus whether high-resolution  stellar velocity dispersion measurements were required.
Repeated observations were therefore done if either the S/N of the broad Balmer lines (\hbeta\ or \halpha) or the S/N and/or spectral resolution of the stellar absorption features were too low.  Repeated observations with higher spectral resolution but limited spectral range were primarily done for obscured AGN (Sy1.9 and Sy2) to measure velocity dispersions (and deduce BH masses), as velocity dispersion measurements are much more difficult and less reliable for AGN-dominated continuum.   We did not reobserve targets with acceptable spectra and measurements from the Sloan Digital Sky survey \citep[SDSS;][]{York:2000:1579} included in Data Release 16 \citep[DR16;][]{Ahumada:2020:3}.

As the goal of BASS is to provide the best and most complete set of derived measurements (e.g. for narrow lines, broad lines, velocity dispersions), we did not require that all studies use the same single best spectra for an AGN in the DR2 as was done in the DR1.  So for instance, a single Sy\,1.9 AGN may have a broad \Halpha\ measurement from SOAR/Goodman using the lower-resolution 400 lines mm$^{-1}$ setting, along with measurements of narrow emission lines, but have a velocity dispersion measurement from CaT using the 1200 line mm$^{-1}$ setting.

 We also did not specifically exclude sources with high Galactic extinction if there was an obvious optical counterpart. This resulted in some spectra with very high extinctions (between $A_\mathrm{V}=5$ and 10 mag), that are only suitable for basic emission-line identification/classification and for redshift measurement.

\subsection{Comparison with BASS DR1}
\label{sec:comp_with_DR1}

The BASS DR1 was composed mostly of past archival optical spectroscopic data from a variety of sources. Almost all the observations were from smaller (1.5-2.5m) telescopes, with only 35 from the Palomar 5m telescope (DBSP) and 29 from the 8.1m Gemini North and South telescopes. Many of the spectra were taken from various surveys and studies that used various reduction routines, leading to substantial inhomogeneity in quality and parameter constraints. Some of the spectra, particularly those from the 6dF Galaxy Survey \citep[6dFGS;][]{Jones:2009:683}, had no proper flux calibration.  Apart from the subset of DR1 spectra taken from the SDSS, most BASS DR1 spectra had a spectral resolution that is too low ($R<1000$) to robustly measure stellar velocity dispersions. Finally, in most cases, the spectral setups did not include coverage below 4000\,\AA\ or beyond 7000\,\AA.

The DR2 results, which are based primarily on Palomar/DBSP or VLT/X-Shooter spectra, are largely a separate release from DR1, even though the AGN samples overlap.  Aside from 142 SDSS spectra commonly used for both samples, the only other spectral overlap between DR1 and DR2 is for 35 Palomar/DBSP observations, which were uniformly reprocessed in DR2 using the new \molecfit\ \citep{Smette:2015:A77} procedure for telluric corrections.  

For the BASS DR2 catalogs of broad and narrow emission line measurements, we allowed some DR1 spectra to be used to have the best and most complete set of derived quantities (i.e., BH mass, etc.).  This was done because the two main DR2 observing setups (VLT/X-Shooter and Palomar/DBSP) used for the majority of sources both had the break in the blue and red CCDs at $\sim$5500\,\AA, making measurements of the spectral complex around the (redshifted) \hbeta\ line problematic (e.g. at $z\approx0.1$).  For stellar velocity dispersion measurements, only DR2 data were used.   

\subsection{NIR DR2}
\label{sed:nir_dr2}
The goal of the BASS NIR spectroscopy is to obtain wide spectral coverage across the full NIR range (${\sim}1-2.4\,\micron$) for a large sample of BAT AGN.  The 102 AGN composing the NIR DR1 \citep{Lamperti:2017:540} were primarily for nearby AGN ($z<0.075$) observed with the SpeX instrument \citep{Rayner:2003:362} at the NASA Infrared Telescope Facility (IRTF) in the northern hemisphere ($0.8-2.4\,\micron$) along with archival Gemini/GNIRS data ($0.8-2.5\,\micron$). 

 The NIR DR2 AGN sample includes 233 new NIR spectra of BASS AGN.  The new DR2 data were obtained from the southern hemisphere and include 168 new VLT/X-Shooter \citep{denBrok_DR2_NIR} and 65 Magellan/FIRE spectra \citep{Ricci_DR2_NIR_Mbh}, both with substantially higher spectral resolutions ($R\sim5000-10000$) than the DR1 data ($R\sim800-1000$).  When including the NIR DR1, the total NIR sample provided as part of DR2 includes NIR spectroscopy of 322 unique AGN\footnote{Four DR2 observations overlap with DR1, while nine DR2 Magellan/FIRE observations overlap with VLT/X-Shooter.}  The NIR catalog measurements include broad, narrow, and  coronal emission line measurements between $1-2\,\micron$, but not the $K$-band ($2-2.5\,\micron$).
 
 The NIR spectroscopic survey of 322 AGN is distinct from the DR2 optical spectroscopy, as it is an interim release rather than a complete sample (i.e., the NIR DR2 is very far from providing NIR spectra for all \NAGN\ AGN in the optical BASS DR2).  Moreover, this NIR survey release also includes some 105-month BAT sources \citep{Oh:2018:4} and the VLT/X-Shooter spectroscopy only includes observations carried out through 2019 October.


Future releases (i.e., the NIR DR3) will include additional NIR spectroscopy efforts within BASS that are currently ongoing, including additional VLT/X-SHOOTER, Magellan/FIRE, and Palomar/TSpec observations, as well as velocity dispersion measurements and emission-line fitting in the $K$ band (${\sim}2-2.4\,\micron$). As of 2021 July, an additional 267 BASS AGN have NIR spectroscopy that are not part of the DR2 release. 
 
\subsection{Survey Uniformity and Completeness}
\label{sec:uniformity}

BASS DR2 is nearly spectroscopically complete with respect to the 70-month BAT all-sky survey, with $>99$\% of the AGN having spectra (i.e., 858/(858 AGN+7 unknown/confused sources)).  There are, however, some additional considerations in terms of uniformity and completeness for the BASS survey, which we briefly mention here and review in more detail in \citet{Koss_DR2_catalog}.  
The 70-month BAT survey reaches a 14--195\,\kev flux sensitivity level of $1.34\times10^{-11}$\,\ergcms\ over 90\% of the sky \citep[assuming a power-law spectral shape fixed to the Crab $\Gamma\sim$2.15;][]{Baumgartner:2013:19} but $1.03\times10^{-11}$\,\ergcms\ over 50\% of the sky, meaning there is some small ($\sim$20\%) variation in sensitivity across the sky. Near the Galactic plane, and particularly in the crowded field of the Galactic center with strong Galactic X-ray background radiation, the source density is considerably lower, and the sensitivity of the BAT survey is a factor of two lower \citep{Markwardt:2005:L77}.

The all-sky BAT survey also suffers from biases against harder and/or highly obscured AGN.  Due to the BAT detection method, which is calibrated to the Crab ($\Gamma=2.15$), AGN with harder intrinsic X-ray spectral energy distributions (SEDs; e.g. $\Gamma=1.7$ vs. $\Gamma=2.1$) will suffer roughly 10\% reduced sensitivity \citep{Koss:2013:L26}.  A more significant completeness correction is needed for highly obscured AGN, where BAT detects 90\% of the flux for an AGN with  $\NH=3\times10^{23}\,\cmii$, 50\% for $\NH=2\times10^{24}\,\cmii$, and $\sim$10\% (or less, depending on models) at $\NH=10^{25}\,\cmii$ \citep[e.g.,][]{Koss:2016:85}.

\section{DR2 Papers, Catalogs, and Key Science Results}
\label{sec:papers_catalogs}

\begin{deluxetable}{llll}
\tablecaption{Summary of BASS DR2 Papers \label{tab:bass_papers}}
\tablehead{
\colhead{BASS \#}&\colhead{Short Title}& \colhead{Major Measurements/Science}& \colhead{Ref.}}
\startdata
XXI&Overview&\mbh\ and \lledd\ vs. other surveys&This paper\\
XXII&Catalog&Counterparts, z, Spectra, Best \mbh&\cite{Koss_DR2_catalog}\\
XXIII&MIR Diagnostic for Absorption&Obscuration in MIR&\cite{Pfeifle:2022:3}\\
XXIV&Narrow line measurements&BPT Diagnostics&\cite{Oh_DR2_NLR}\\
XXV&Broad-line measurements&Sy\,1-Sy\,1.9 BLR vs. \NH, Broad lines&\cite{Mejia_Broadlines}\\
XXVI&Stellar Velocity Dispersions&Sy\,1.9 and Sy\,2 \sigs\ \mbh&\cite{Koss_DR2_sigs}\\
XXVIII&NIR High-Ionization and Broad Lines&NIR spectra, NIR lines&\cite{denBrok_DR2_NIR}\\
XXIX&NIR view of the BLR Effects of Obscuration&NIR BLR vs. \NH&\cite{Ricci_DR2_NIR_Mbh}\\
XXX&Distribution Function of Eddington Ratios&XLF, ERDF, BHMF&\cite{Ananna_DR2_XLF_BHMF_ERDF}\\
\enddata
\tablecomments{List of all papers in the BASS DR2 release and associated major measurements and science.  BASS XXVII was published separately from the DR2 release.}
\end{deluxetable}

The DR2 papers \autoref{tab:bass_papers} provide a combination of large datasets of measurements and derived quantities, and scientific results for either the entire DR2 sample or considerable subsets of data.  We provide a short review of these papers here, to explain the different measurements and catalogs provided in the various papers and to highlight key scientific findings.  
In some papers (e.g., narrow emission line measurements) a very wide spectral range was critical (3200--10000\,\AA), while in others (e.g., velocity dispersions) high spectral resolutions were critical; therefore, sometimes spectra with high spectral resolution but narrow wavelength range were used (e.g., for the calcium triplet, CaT, 8498, 8542, and 8662\,\AA). 
All DR2 spectra will be provided at the BASS website.\footnote{\url{http://www.bass-survey.com}}

The main, detailed DR2 catalog paper \citep[BASS XXII,][]{Koss_DR2_catalog} gives an updated list of counterparts and a complete summary of observing characteristics and reduction procedures for the \Nspec\ optical spectra of the \NAGN\ AGN from the BAT 70-month sample.  It discusses various issues for basic AGN observational studies (source confusion,  chance alignment of multiple AGN leading to flux boosting, dual AGN, etc.), including a revised list of beamed and lensed AGN identifications.  This paper also provides a set of overall best-derived measurements (e.g., redshifts and distances, bolometric luminosities, and \mbh) since multiple BH mass estimates are available for some sources from various tracers (i.e., different broad lines, velocity dispersions, and/or high-quality literature measurements).  The overall best available measurements from this catalog are then used consistently in subsequent scientific papers.

Narrow optical emission line measurements over the range 3200--10000\,\AA\ (e.g., \NeV, \hbeta, \OIII, \halpha, \NII, \SIII) are presented in BASS XXIV \citep[][]{Oh_DR2_NLR}.  In that paper, we employ a full-range spectral fitting procedure, incorporating both stellar population synthesis models and empirical stellar templates, which deblends complex nebular emission-line features from the stellar components. We also study AGN subtypes as a function of X-ray column density; strong-line ratio diagnostic diagrams (BPT) and their links to Eddington ratio; and line width comparisons between X-ray BAT AGN and optical SDSS AGN.  


Broad emission line measurements (\halpha, \hbeta, \MgII, and \CIV) and derived quantities are the focus of BASS XXV \citep[][]{Mejia_Broadlines}.  This paper includes virial estimates of BH mass (\Mbh) which are used as the best black hole mass measurement throughout DR2 and also allow estimates of the Eddington ratio (\lledd).  The use of \MgII\ and \CIV\ is reserved for the beamed AGN with discernible broad lines, which are at higher redshifts (e.g., $z\gtrsim1$), where rest-frame \hbeta\  falls outside of the optical range. This paper concludes that the innermost part of the broad-line region (BLR), which contributes the highest velocity emission, is preferentially absorbed in obscured AGN ($\logNH> 22$) and/or Sy1.9. This leads to a significant underestimation of the line flux compared to unobscured sources, which then strongly underestimates BH mass. These discrepancies typically exceed 1 dex and may reach 2 dex for heavily obscured AGN ($\logNH\geq 24$). We provide some prescriptions for corrections.

Central velocity dispersion measurements (e.g., from \Cahk, \mgb, or CaT regions) of obscured systems (Sy\,1.9 and Sy\,2) are investigated in BASS XXVI \citep[][]{Koss_DR2_sigs}.  This paper finds that BASS AGN have much higher velocity dispersions than the more numerous optically selected narrow-line AGN (i.e., $\approx$150 vs. $\approx$100\,\kmps), but also that BASS AGN are {\it not} biased toward the highest velocity dispersions seen in massive ellipticals (i.e., $>$250\,\kmps).   Additionally, despite sufficient spectral resolution to resolve the velocity dispersions associated with the bulges of relatively small SMBHs ($\sim10^{4}-10^{5}$\,\Msun), we do not find a significant population of such AGN, which (given the BAT flux limit) would have presented super-Eddington accretion rates.

\citet{Ananna_DR2_XLF_BHMF_ERDF} use the highly complete set of BASS DR2 measurements to derive the intrinsic XLF, BH mass function (BHMF) and Eddington ratio distribution function (ERDF), for both obscured and unobscured low-redshift AGN using the BAT sample. It employs an elaborate forward-modeling approach to derive the intrinsic XLF, BHMF and BHMF from the observed distributions, while accounting for various selection effects. 
We find that the intrinsic ERDF of narrow-line (Type~2) AGN is significantly skewed toward lower Eddington ratios than that of broad-line (Type~1) AGN, while the BHMFs of these subsamples are consistent with each other.
This result offers insights into the geometric structure of the obscuring ``torus''  and lends support to the radiation-regulated unification scenario \citep{Ricci:2017:488}, which suggests that  radiation pressure dictates the geometry of the dusty obscuring structure around an AGN. 
The XLF, BHMF, and ERDF are also used to investigate the AGN duty cycle in the low-redshift universe.



\citet{Pfeifle:2022:3} investigate the relationship between MIR colors and X-ray column density. Heavily obscured BAT AGN are found to be more heavily X-ray suppressed, displaying lower ratios of \Lsoftobs$/ L_{\mathrm{12\,\micron}}$, and they display ``redder'' MIR colors compared to unobscured AGN. This paper develops diagnostic criteria that are designed to select both highly complete and highly reliable samples of heavily obscured AGN ($\logNH >23.5$). We also derive expressions relating the luminosity ratios and column density, to predict the AGN column density in lower count-rate X-ray SEDs, where detailed spectral modeling is impossible. 
These diagnostics could be used on future samples of AGN, such as those being discovered by eROSITA \citep{2021A&A...647A...1P}, to efficiently distinguish between heavily obscured and unobscured AGN.

\citet{denBrok_DR2_NIR} provide a detailed analysis of the NIR coronal lines (CLs, ionization potential $\chi>100$ eV) and test their usage as indicators of AGN activity by comparing their strength, in particular that of \SiVI, to the X-ray flux. A key finding is that CLs correlate more tightly (i.e., smaller scatter) with the X-ray fluxes than with the optical \OIII\ line fluxes. Even in these bright AGN, in only about half of the sources is a CL detected, limiting the extent to which CLs can be used as tracers of AGN activity. This study finds a clear trend of line blueshifts with increasing ionization potential in several CLs, such as \SiVI, \SiX, \SVIII, and \SIX, which elucidates the radial structure of the CL region.

Finally, \citet{Ricci_DR2_NIR_Mbh} investigate the NIR BLR using Pa $\alpha$, Pa $\beta$, and \HeIir, and associated virial BH mass estimates. The NIR regime is less affected by dust than the optical and can thus trace the innermost and fastest-moving BLR gas --- even in the presence of mild obscuration.  The study finds that the velocities of the BLR gas as estimated from the FWHMs of \halpha\ and the NIR lines in Sy 1--1.9 agree and are independent of the level of BLR extinction or obscuration (for $\logNH<23.5$), but the broad line luminosities are suppressed with increasing obscuration, biasing virial-based \mbh\ estimates. 
The latter finding is in agreement with the conclusion of \cite[][see above]{Mejia_Broadlines}.
The line luminosity decrement and the obscuration level at which it occurs change as a function of wavelength, with \Halpha\ experiencing a higher decrement than Pa$\alpha$ (above $\logNH\simeq21.0$ and $21.9$, respectively). Thus, we caution against relying solely on \Halpha-based single-epoch BH mass estimates when $\logNH\gtrsim21$, and on NIR lines when $\logNH\simeq22$. 
A less biased proxy for the BLR radius in virial-based \mbh\ should be used at higher \NH, such as $L_X$. 


\section{Other BASS Observing Campaigns}
\label{sec:other_obs}

Beyond the complete coverage with optical spectroscopy and the extensive NIR spectroscopy that are the main components of the BASS DR2, the BASS project aims to obtain and analyze large multiwavelength data sets for the BAT AGN, in the X-ray, UV, optical, IR, FIR/submillimeter, and radio regimes. 
These include ongoing legacy BASS campaigns, past observations through BASS and the community, and various all-sky surveys (e.g. GALEX, GAIA, 2MASS, WISE,  Akari, IRAS). The status of targeted observational programs for the 70-month AGN as of 2021 July is summarized in \autoref{tab:bass_other} and on the BASS website.\footnote{\url{http://www.bass-survey.com}}  

In addition to this, partial sky coverage exists from several wide-field surveys for hundreds of AGN.   Multiband high-quality optical imaging ($<$2\arcsec) exists for the majority ($>$80\%) of BASS AGN from the SDSS, the Panoramic Survey Telescope and Rapid Response System \citep[Pan-STARRS][]{Chambers:2019}, the DESI Legacy Imaging Surveys \citep{Dey:2019:168}, The VLT Survey Telescope ATLAS \citep{Shanks:2015:4238}, and targeted studies like \citet{Koss:2011:57}.  In the NIR, coverage exists in the VISTA Hemisphere Survey \citep[VHS;][]{McMahon:2013:35} and the UKIRT Hemisphere Survey \citep[UHS;][]{Dye:2018:5113}. In the radio coverage at 1.4 GHz exists for more than half of the BAT AGN \citep{Wong:2016:1588} from the Faint Images of the Radio Sky at Twenty Centimeters \citep[FIRST;][]{Becker:1995:559} and NRAO VLA Sky survey \citep[NVSS;][]{Condon:1998:1693}. 

The BAT AGN have the largest coverage in the X-rays and UV, with all \NAGN\ observed with Swift/XRT and also observed with Swift/UVOT in the UV and optical. This results in a vast database of simultaneous X-ray and UV observations, obtained since the launch of Swift in 2004.  For instance, for broad-line (Sy1) AGN there are 32,184 distinct observations, due to the slewing nature of Swift and their intensive coverage in legacy observations of BAT AGN.  

As the brightest ultra-hard-X-ray-selected AGN in the sky, many archival observations are available from several other X-ray telescopes.  In particular, many BAT AGN have also been observed as part of 20 ks filler observations in the NuSTAR BAT Legacy Survey,\footnote{\url{https://www.nustar.caltech.edu/page/legacy\_surveys}} which continues to observe approximately six BASS sources each month.  A Chandra Cool Target program,\footnote{Chandra-BASS (C-BASS); \url{https://cxc.harvard.edu/target_lists/CCTS.html}} which started in 2019 January, is also observing nearby BAT AGN ($z<0.1$).

Another subset of observing programs focuses on high spatial resolution imaging ($\sim$100 pc) that can be achieved for nearby BASS AGN ($z<0.1$).  This includes a recent HST SNAP program with the Advanced Camera for Surveys (ACS), which has obtained $i$-band (F814W) images for 154 DR1 BAT AGN at $z<0.1$ \citep{2021ApJS..256...40K}.  A large HST SNAP program, approved through 2022, aims to obtain near-UV (${<}$3000\,\AA) imaging of BASS AGN, followed up with simultaneous X-ray and UV/optical observations of the nuclear AGN emission with Swift and ground-based optical imaging in $griz$.  \citet{Koss:2018:214a} published 98 Keck/NIRC2 AO-assisted NIR observations (in the $Kp$ band) for a volume-limited sample of AGN ($z<0.1$), along with many archival HST NIR observations of more nearby galaxies available from earlier HST/NICMOS studies \citep[e.g.][]{Hunt:2004:707}.  Finally, approved high-resolution AO NIR imaging and spectroscopy are also underway through 2022 using Keck/NIRC2, Keck/OSIRIS, and Gemini/GSAOI, with a focus on hidden galaxy mergers and dual, small-separation AGN in obscured systems.

Finally, another set of survey programs are broadly focused on connecting the key star-formation-related properties of the AGN hosts, such as SFR and molecular and atomic gas, with AGN activity in nearby AGN, using high-resolution and high-sensitivity observations over the IR-millimeter-radio spectral regimes.  These programs include earlier studies carried out with Spitzer \citep[e.g.,][]{Weaver:2010:1151} and Herschel \citep[e.g.,][]{Mushotzky:2014:L34}, which focused on nearby BAT AGN and were then followed up in the submillimeter and radio.  
The Herschel program to measure star formation observed 317 of the nearest BAT AGN ($z<0.05$), which were later followed up with more recent measurements of host galaxy molecular gas using CO lines \citep{Koss:2021:29}.  
and 22 GHz \citep[e.g.,][]{Smith:2020:4216} observations with the JVLA. A yet nearer-distance sample ($z<0.025$) is being targeted for HI mapping using the JVLA.  More molecular gas observations using APEX have also been approved for these sources, through 2022.  
A program to obtain 100 pc resolution CO(2-1) measurements using ALMA was done for 33 nearby and luminous AGN.  High spatial resolution radio observations (0.2--0.5 pc resolution) that form a complete volume-limited sample out to 40 Mpc for AGN above \Lbat$>10^{42}$ \ergps, has also been done for a sample of 37 objects using C-band Very Long Baseline Array (VLBA).

\begin{deluxetable}{lcrclc}
\tablewidth{0pt}
\tablecaption{Summary of BASS DR2 Data \label{tab:bass_dr2}}
\tablehead{
\colhead{Telescope}& \colhead{Instrument}& \colhead{Total}& \colhead{Range (\AA)}& \colhead{Slit Width (\arcsec)}& \colhead{$R$}\\\colhead{(1)}& \colhead{(2)}& \colhead{(3)}& \colhead{(4)}&\colhead{(5)}& \colhead{(6)}}
\startdata
Palomar&DBSP&502&3150-10500&1.5&1220/1730\\
VLT&X-Shooter&233&2990-10200&1.6/1.5&3850/6000\\
APO&SDSS&177&3830-9180&3&1760/2490\\
du Pont&BC&119&3000-9070&1&480\\
Archival\tablenotemark{a}&Various&90&Various&Various&Various\\
VLT&FORS2&70&3400-6100&1&830\\
SOAR&Goodman&67&4560-8690&1.2&890/1630\\
Keck&LRIS&21&3200-10280&1&1280/1810\\
Magellan&MAGE&12&3300-10010&1&3850\\
VLT&MUSE&6&4800-9300&2&1850/3150\\
\hline \hline Velocity Dispersion Setups\tablenotemark{b}\\ \hline \hline
SOAR&Goodman&86&7900-9070&1.2&4720\\
Palomar&DBSP&66&3970-5499/8050-9600&2&2170/4720\\
\hline \hline NIR DR2\\ \hline \hline
VLT&X-SHOOTER&168&10240-24800&0.9&5400\\
Magellan&FIRE&65&8000-25000&0.6&6000\\
\enddata
\tablecomments{Column (1): telescope. Column (2): instrument. Column (3): total number of DR2 spectra observed with this setup. Column (4): wavelength range for the most common setup with the telescope. Column (5--6): slit width and resolving power for the most common setup.  In some cases larger or smaller slit widths (e.g., 1\farcs{5}\ vs. 2\arcsec) were used resulting in different resolutions.  See \citet{Koss_DR2_catalog} for a detailed list of instrument setups.  Two values are listed when the instrument had both a blue and red arm with different settings. Resolving power is wavelength dependent in some cases and so the values are given at 5000\,\AA\ and 8500\,\AA\ depending on the spectral range.}
\tablenotemark{a}{The archival sample is from earlier surveys that were not included in the DR1, including ROSAT AGN that overlap with BASS in unpublished or published \citep{Grupe:2004:156} works, from the Palermo surveys of Swift BAT AGN \citep[][]{Rojas:2017:A124}, or as part of an atlas of low redshift AGN \citep{Ho:2009:398}. While not typically used in catalog measurements because of new DR2 data, we include these spectra for long-term studies of changing-look AGN \citep[e.g.,][]{MacLeod:2019:8}.}
\tablenotemark{b}{These setups were done for velocity dispersion measurements of obscured AGN (e.g., Sy1.9 and Sy 2).}
\end{deluxetable}

\begin{deluxetable}{lcccl}
\tabletypesize{\small}
\tablewidth{0pt}
\tablecaption{Additional BASS and Archival Multiwavelength Data  \label{tab:bass_other}}
\tablehead{
\colhead{Data Set / Telescope}& \colhead{Spectral Bands}& \colhead{$N_{\rm AGN}$}& \colhead{Focus}& \colhead{Investigators}
\\
\colhead{(1)}&\colhead{(2)}& \colhead{(3)}& \colhead{(4)}&\colhead{(5)}
}
\startdata
NuSTAR & 3-70 keV & 527 & & Ricci, Koss, Archival\\
Swift XRT & 0.5-10 keV & 858 & & Ricci, Koss, Archival\\
XMM-Newton & 0.5-10 keV & 386 & & Ricci, Koss, Archival\\
Chandra & 0.5-8 keV & 384 & & Koss et al., Archival\\
Suzaku & 0.3-10 keV & 210 & & Archival\\
HST & F225W & 54+ & Sy1, $z{<}0.1$ & Koss et al.\\
Swift UVOT & UV ($W2$, $M1$, $W1$)/ $UBV$ & 812 & & Ricci, Koss, Archival\\
XMM OM & UV ($W2$, $M1$, $W1$)/ $UBV$ & 342 & & Ricci, Koss, Archival\\
SNIFS IFU & 3200-10200 \AA & 46 & $z{<}0.05$ & Koss et al.\\
MUSE IFU & 4800-9300 \AA & 84 & & Archival\\
HST & F606W & 157 &  & Archival\\
HST & F814W & 243 & $z{<}0.1$ & Barth, Archival\\
HST & F160W & 104 & & Archival\\
NIR AO (Keck/NIRC2, Gemini/GSAOI) & $H$, $K$ & 98+ & $z{<}0.1$ & Koss, Treister et al.\\
NIR AO IFU (Keck/OSIRIS, VLT/SINFONI)& $H$, $K$ & 108+ & & Koss et al.\\
VLT/VISIR AO & 8-13\,\micron & 125 & $z{<}0.01$ & Asmus et al.\\
Spitzer IRS low res. & 5.3-35\,\micron & 175 & & Archival\\
Spitzer IRS high res. & 10-37\,\micron & 140 & $z{<}0.05$ & Weaver et al.\\
Herschel&70, 160, 250, 350, 500\,\micron & 317 & $z{<}0.05$ & Mushotzky, Shimizu et al.\\
JCMT/Scuba 2&450, 850\,\micron & 63 & $z{<}0.05$ & Koss et al.\\
ALMA & 100 GHz &99 & & Archival\\
APEX/IRAM/JCMT&CO 1-0/CO 2-1&305+&z$<$0.05&Koss, Shimizu, et al.\\
ALMA&CO 1-0/CO 2-1 & 156 & $<$100 Mpc & Izumi et al., Archival\\
(J)VLA&22 GHz&232 & $z{<}0.05$ & Smith, Mushotzky, et al.\\
VLBA & 5 GHz & 37 & $<$40 Mpc & Secrest et al.\\
GBT & HI & 96 & $z{<}0.05$ & Winter et al.\\
ATCA/(J)VLA/WSRT/GMRT&HI mapping & 98 & $<$120 Mpc & Chung, Wong, et al.\\
\enddata
\tablecomments{Column 
(1): telescope or instrument for the survey data. If the data are substantially similar (e.g., AO imaging in the same filter), we have grouped telescopes. 
Column (2): wavelength, frequency, line, filter, or energy band. 
Column (3): total number of unique AGN from the Swift BAT 70-month catalog, which includes a total of \NAGN\ AGN. This number does not include 105-month AGN which will be released in subsequent catalogs (BASS DRs).  The ``+'' sign indicates approved and/or ongoing additional observations. 
Column (4): indicates whether (part of) the observations were focused on a volume-limited sample, and/or particular AGN subxclass.  
Column (5): main investigators for survey data. ``Archival'' indicates that the majority of corresponding data are from disparate observing programs.}
\end{deluxetable}



\section{Overall Survey Results}
\label{sec:ovearall_results}

The BASS survey is a spectroscopically complete for 100\% (\NAGN/\NAGN) of the AGN identified in the 70-month BAT all-sky survey outside of 7 sources deep within the Galactic plane ($\lvert b\rvert{<}3^{\circ}$) which we were unable to target.  The BASS DR2 reports redshifts for 99.9\% (857/\NAGN) of the AGN, excluding only one continuum dominated blazar with a foreground Galactic star.  This includes  \Nnewz\ redshifts reported for the first time.  Outside of the Galactic plane ($\lvert b\rvert{>}10^{\circ}$) the survey completeness in BH mass measurements from broad lines or stellar velocity dispersion is $98\%$ for all unbeamed AGN because of the typically lower extinction in these regions.  The remaining sources without BH mass measurements are mostly double-peaked/and or assymetric broad-line AGN and high redshift Sy 2 ($z>0.1$), where high-quality velocity dispersion measurements are difficult. For beamed AGN we report black hole masses only for the BZQ class based on broad lines; for continuum-dominated blazars (BZB) velocity dispersion measurements are difficult to the AGN-dominated continuum.

\subsection{Survey Completeness and Measurables}
\label{sec:completeness}

In this section, we briefly illustrate the distributions of some of the key AGN-related properties of the BAT 70-month AGN, as determined through the BASS DR2 measurements We also briefly discuss the reliability and limitations of the measurements.  A more elaborate forward-modeling approach accounting for various survey selection effects associated with a flux-limited survey with various levels of obscuration is described in \citet{Ananna_DR2_XLF_BHMF_ERDF}.

\begin{figure*} 
\centering
\includegraphics[width=\textwidth]{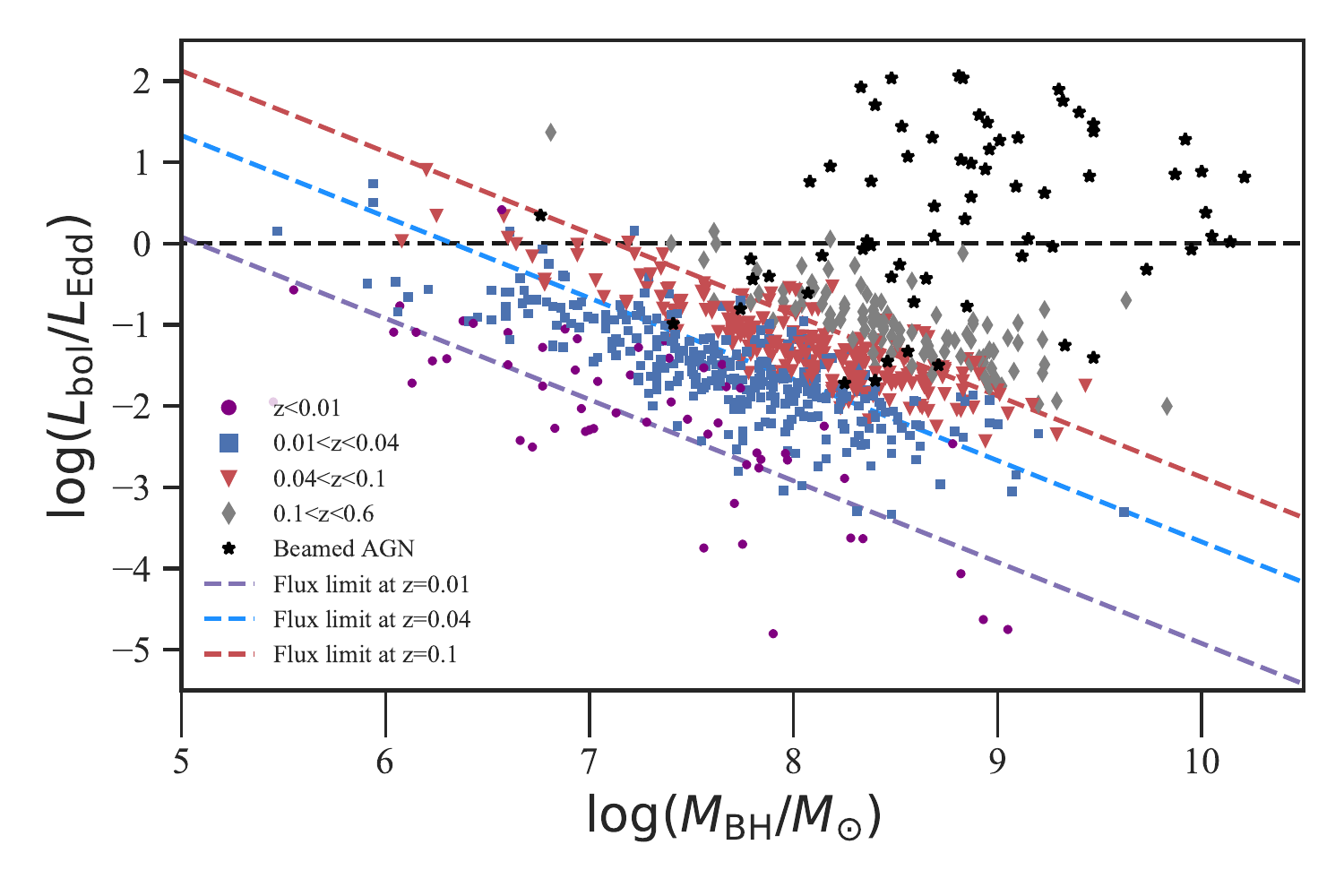}
\caption{Distribution of BH masses (\mbh) and Eddington ratios (\lledd) for the entire sample of BASS AGNs for which redshift measurements are available, including unbeamed AGN (e.g., Sy 1, Sy 1.9, and Sy 2) and beamed AGN (BZQs).  The lower limits on \mbh\ and \lledd\ due to the survey flux limits are illustrated with dashed (antidiagonal) lines, for various redshifts corresponding to $z{=}0.01$ (\lbol${=}1.8{\times}10^{43}$\ergps), $z{=}0.04$ (\lbol${=}3.2{\times}10^{44}$\ergps), and $z{=}0.1$ (\lbol${=}2.0{\times}10^{45}$\ergps).   Errors in \mbh\ are of order 0.5 dex owing to systematic uncertainties in virial and $\sigma_*$-based scaling relations \citep[e.g.,][]{Ricci_DR2_NIR_Mbh}.  The BASS unbeamed AGN occupy narrow, redshift-dependent slices of the $\mbh-\lledd$ plane due to the Eddington limit and survey flux limits (see text for details).  Interestingly, we find that the $\log$(\lledd) tends to be bounded  at -2.5 to -3 possibly associated with BAT identifying disk accretion primarily rather than inefficient accretion and the upper bounds at the Eddington limit (except for beamed AGN).  The lower bound of the BH mass distribution corresponds to \lmbh=5, where the range pushes into intermediate-mass BHs and would only be sensitive to Eddington/super-Eddington accretion, if it exists.  The upper limit at \lmbh=10 is largely owing to the space density of massive BHs.   }
\label{fig:mbh_lledd_diff_z}
\end{figure*}

Summaries of the typical bolometric luminosities, BH masses, Eddington ratios, and line-of-sight hydrogen column densities are provided in \autoref{tab:median_unbeamedagn} and \autoref{tab:median_beamedagn} for unbeamed and beamed AGN, respectively.  
Their BH mass vs. Eddington ratio parameter space is shown in \autoref{fig:mbh_lledd_diff_z}. Overall, the completeness quite similar for Sy\,1, Sy\,1.9, and Sy\,2 ($\sim98\%$) outside the Galactic plane. 

As a survey of the nearest and brightest hard X-ray-selected AGN in the sky, the measurement-associated uncertainties on BASS measurements are typically small.  Apart from the beamed sources (e.g., blazars or ``BZB'') that lack emission lines and a handful of sources that are located in extremely crowded (Galactic plane) regions, the spectra of all BASS sources have multiple emission lines for robust redshift measurements.  The uncertainties on BH mass determinations from \sigs\ are dominated by the systematics on the \mbh-\sigs\ scaling relation \citep[e.g., $\sim$0.3-0.5 dex,][]{Marsden:2020:61a} rather than on the scatter found in repeat observations \citep[$\sim$0.1-0.2 dex]{Koss_DR2_sigs}.  Similarly, the uncertainties on BH masses derived through spectral analysis of broad lines may reach $\sim$0.3-0.4 dex \citep{Peterson:2014:253} whereas the measurement uncertainty is much lower \citep[i.e., 0.1 dex][]{Mejia_Broadlines}.   
The uncertainties on \NH\ are $\sim$0.05 dex for \logNH$<$23.5 and $\sim$0.3 for \logNH$>$23.5 \citep{Ricci:2017:17}.
For intrinsic X-ray luminosity, the errors are typically $<0.1$ dex \citep{Lanz:2019:26}, unless the AGN are CT, in which case the typical errors can reach $\approx0.4$ dex \citep{Ricci:2015:L13}.  The bolometric luminosities are calculated from the intrinsic luminosities in the 14--150\,keV range as shown in \citet[see their Table 12]{Ricci:2017:17}, using a bolometric correction of 8 \citep[see, e.g.,][]{Koss_DR2_catalog}. In this case, the uncertainties are in the range of 0.2 dex \citep[e.g.,][]{Trakhtenbrot:2017:800}.

\begin{figure*} 
\centering
\includegraphics[width=12cm]{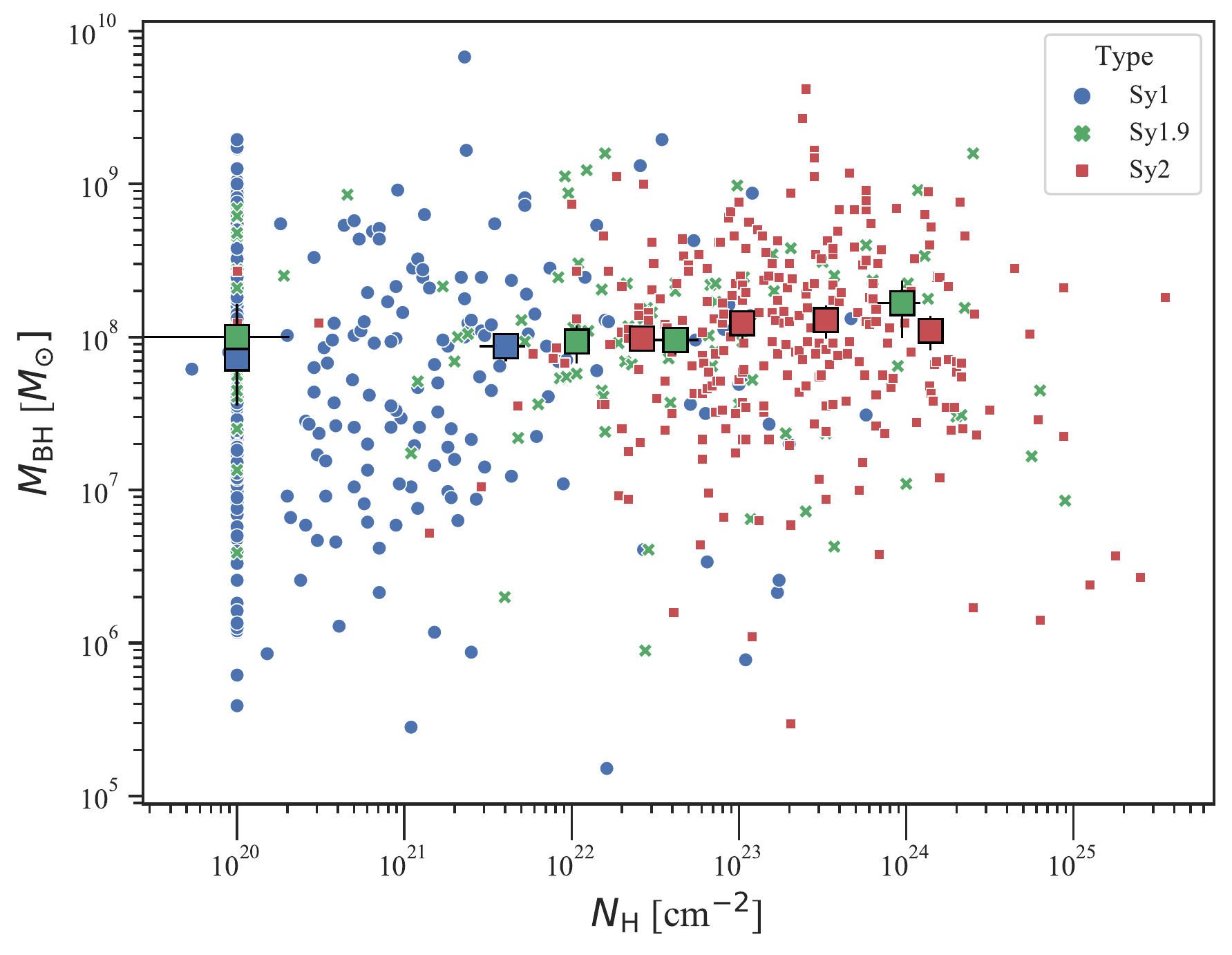}
\includegraphics[width=12cm]{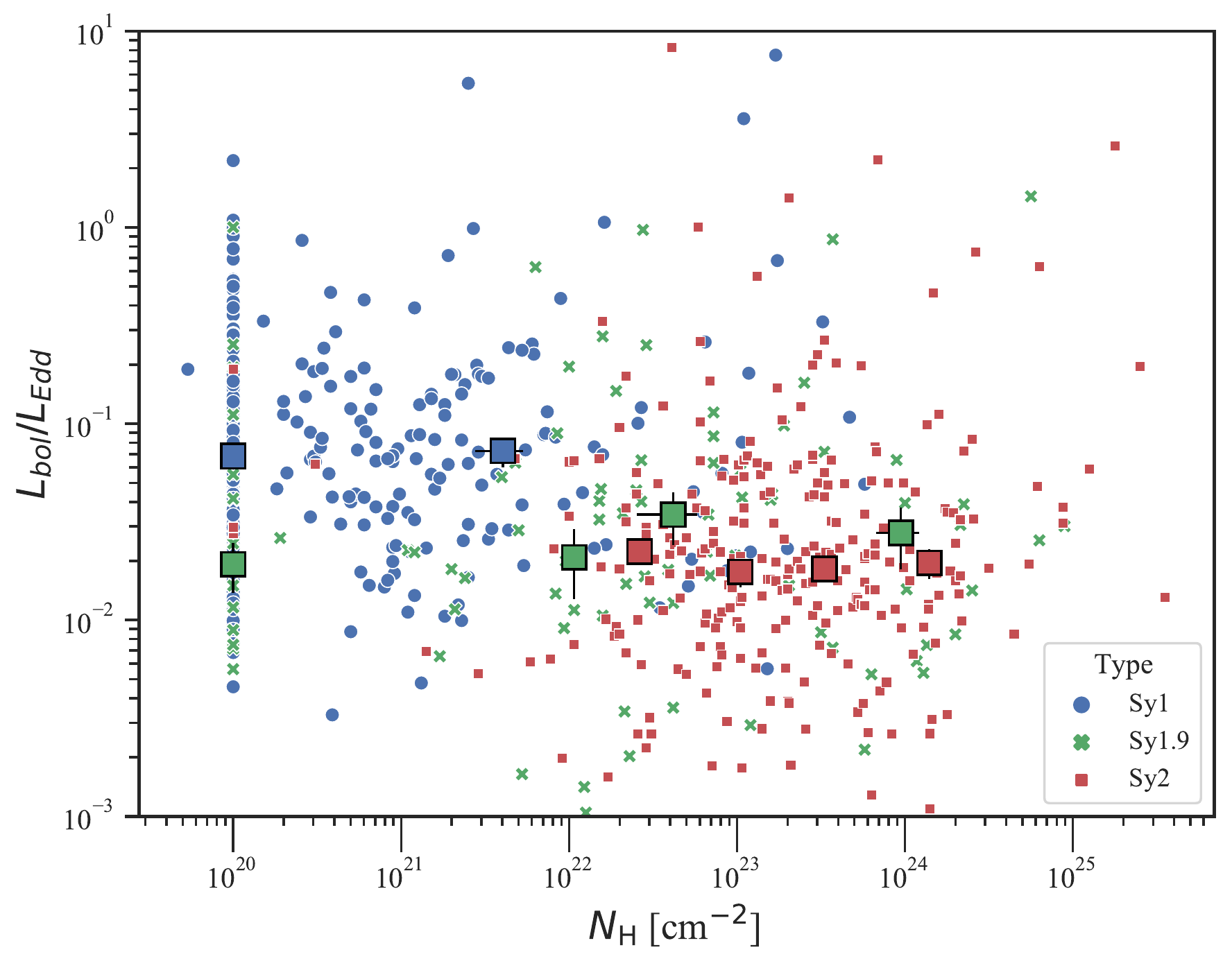}
\caption{Summary scatter plots of the line-of-sight hydrogen column density (\NH, an obscuration indicator) vs. BH mass (top) and Eddington ratio (bottom). Different symbols mark subclasses of unbeamed BASS AGN (e.g., Sy 1, Sy1.9, and Sy 2).  
X-ray measurements of $\NH=10^{20}\,\cmii$ are essentially upper limits and represent sources with no sign of obscuration.  The large squares indicate the binned medians for each subclass.  Error bars on the plotted median values are equivalent to 1$\sigma$ and calculated based on a bootstrap procedure with 100 realizations. The bin sizes were constructed to have equal numbers of sources in each bin.  The Sy 1 AGN tend to have higher Eddington ratios than the narrow-line AGN (Sy 1.9 and Sy 2). Typical 90\% errors in \NH\ are $<0.2$ dex based on X-ray modeling \citep{Ricci:2017:17}, but higher for heavily obscured AGN log \logNH$>24.5$.}
\label{fig:mbhnh}
\end{figure*}

We also looked at relationships between the line-of-sight column density, as measured from X-ray data, and BH mass and Eddington ratio, as determined from optical spectroscopy (\autoref{fig:mbhnh}).  One clear takeaway is that unobscured (Sy 1) AGNs occupy a region of higher Eddington ratios compared to obscured (Sy 1.9s and 2) AGN.  This echoes the previous BASS DR1-based finding by \citet{Ricci:2017:488}, which supports a scenario where radiation pressure is the main driver of the geometry of the (dusty) circumnuclear gas.  There are no obvious trends between \NH\ and either \mbh\ or \lledd\ within each AGN optical subclass.

\begin{deluxetable}{ccccccccccccc}
\tabletypesize{\footnotesize}
\tablewidth{0pt}
\tablecaption{Summary of Unbeamed AGN Properties \label{tab:median_unbeamedagn}}
\tablehead{
\colhead{Type}& \colhead{$N$}&\colhead{$N$}& \colhead{$z$}& \colhead{$N_{\mbh}$}& \colhead{$N_{\mbh}$}& \colhead{\% Meas \mbh}& \colhead{\% Meas}& \colhead{$\log$ \mbh}& \colhead{$\log$ \Lbol}& \colhead{$\log$ \lledd}& \colhead{$\log$ \NH}\\
\colhead{}& \colhead{}&\colhead{$\lvert b\rvert>10^{\circ}$}& \colhead{}& \colhead{}& \colhead{$\lvert b\rvert>10^{\circ}$}& \colhead{}& \colhead{$\lvert b\rvert>10^{\circ}$}& \colhead{(\Msun)}& \colhead{(\ergps)}& \colhead{}& \colhead{(\nhunit)}
\\
\colhead{(1)}& \colhead{(2)}& \colhead{(3)}& \colhead{(4)}&\colhead{(5)}& \colhead{(6)}& \colhead{(7)}&
\colhead{(8)}&
\colhead{(9)}&
\colhead{(10)}&
\colhead{(11)}&
\colhead{(12)}
}
\startdata
Sy1&359&318&0.050$\pm$0.003&350&311&97&98&7.81$\pm$0.04&44.87$\pm$0.04&-1.17$\pm$0.03&20.0$\pm$0.05\\
Sy1.9&101&86&0.030$\pm$0.004&97&84&96&98&7.98$\pm$0.06&44.59$\pm$0.08&-1.61$\pm$0.09&22.28$\pm$0.13\\
Sy2&292&259&0.029$\pm$0.003&275&253&94&98&8.06$\pm$0.04&44.50$\pm$0.04&-1.71$\pm$0.04&23.27$\pm$0.05\\
\hline
Total&752&663&0.038$\pm0.002$&722&648&96&98&7.96$\pm$0.03&44.67$\pm$0.03&-1.42$\pm$0.03&21.98$\pm$0.06
\enddata
\tablecomments{Summary of the medians and standard error of the median for different populations of unbeamed AGN.  Column (1): AGN optical type based on the presence of broad \hbeta\ and \halpha.  Column (2): total for the whole sample. Column (3): total excluding the Galactic plane region $\lvert b\rvert<10^{\circ}$ where high optical extinction makes measurements more difficult. Column (4): median redshift from optical lines.  Columns (5--8): number of unique AGN with \mbh\ measurements and excluding the Galactic plane region $\lvert b\rvert<10^{\circ}$ where high optical extinction makes measurements more difficult; also listed as percentages.  Columns (9-12): median \mbh, \lbol, \lledd, and \logNH\ for the sample.  }
\end{deluxetable}

\begin{deluxetable}{ccccccccccc}
\tabletypesize{\small}
\tablewidth{0pt}
\tablecaption{Summary of Beamed AGN Properties \label{tab:median_beamedagn}}
\tablehead{
\colhead{Type}& \colhead{$N$}&\colhead{$N$, $\lvert b\rvert>10^{\circ}$}& \colhead{$z$}& \colhead{$N_{\mbh}$}& \colhead{\% Meas \mbh}&  \colhead{$\log$ \mbh}& \colhead{$\log$ \Lbol}& \colhead{$\log$ \lledd}& \colhead{$\log$ \NH}
\\
\colhead{(1)}& \colhead{(2)}& \colhead{(3)}& \colhead{(4)}&\colhead{(5)}& \colhead{(6)}& \colhead{(7)}&
\colhead{(8)}&
\colhead{(9)}&
\colhead{(10)}
}
\startdata
BZQ&74&63&0.88$\pm$0.12&67&91&8.83$\pm$0.09&47.66$\pm$0.16&0.38$\pm$0.12&20$\pm$0.11\\
BZB&22&18&0.13$\pm$0.02&&&&45.81$\pm$0.12&&20.57$\pm$0.12\\
BZG&8&6&0.07$\pm$0.02&&&&45.11$\pm$0.20&&20.81$\pm$0.11\\
Sy1/lensed&1&1&0.65&1&100&8.79&47.18&0.21&20\\
BZQ/lensed&1&0&2.51&&&&49.49&&22.77\\
\hline
Total&106&88&0.33$\pm$0.10&68&&8.83$\pm$0.09&46.53$\pm$0.14&0.38$\pm$0.12&20.54$\pm$0.08\\
\enddata
\tablecomments{Summary of the medians and standard error of the median for different populations of beamed and/or lensed AGN.  Column (1): AGN optical type based on presence of broad lines (BZQ), only host galaxy features lacking broad lines (BZG), or traditional continuum dominated blazars with no emission lines (BZB), or lensing. Column (2): Total for the whole sample. Column (3) total excluding the Galactic plane region $\lvert b\rvert<10^{\circ}$ where high optical extinction makes measurements more difficult. Colun (4): median redshift from optical lines.  Column (5--6): number of unique AGN with \mbh\ measurements and percentages.  Column (7--10): median \mbh, \lbol, \lledd, and, \logNH\ for the sample. }
\end{deluxetable}

\subsection{Redshift Survey Biases}
\label{sec:biases}

Looking more carefully into \autoref{fig:mbh_lledd_diff_z}, the unbeamed BASS AGN typically occupy narrow regions in the BH mass---Eddington ratio plane, in different redshift intervals. 
These could be easily understood as a combination of survey (flux) limits, AGN physics, and demographics. 
For example, the most distant AGN in our sample ($z{>}0.1$) include almost no sources with small BHs ($\mbh{\lesssim}10^7$ \Msun) because these would have to be super-Eddington to be detected.\footnote{That is, $\Lbol\propto \mbh \times \lledd$ for a high-$z$ source would result in a low flux.}   
Likewise, the lowest Eddington ratios sources ($\lledd{\lesssim}10^{-3}$) are also almost exclusively at the highest-mass and lowest-redshift ($z{<}0.01$) systems.  
However, while our sample is affected by strong low-mass and low accretion rate biases,  there are no biases against super-Eddington non-beamed AGN. Notably, such sources are extremely rare in the survey, suggesting that the Eddington limit remains meaningful despite the simplifications in its derivation.\footnote{The alternative is that super-Eddington accretion in SMBHs is extremely X-ray weak \citep[e.g.,][]{2022A&A...657A..57L}.}

\begin{figure*} 
\centering
\includegraphics[width=12cm]{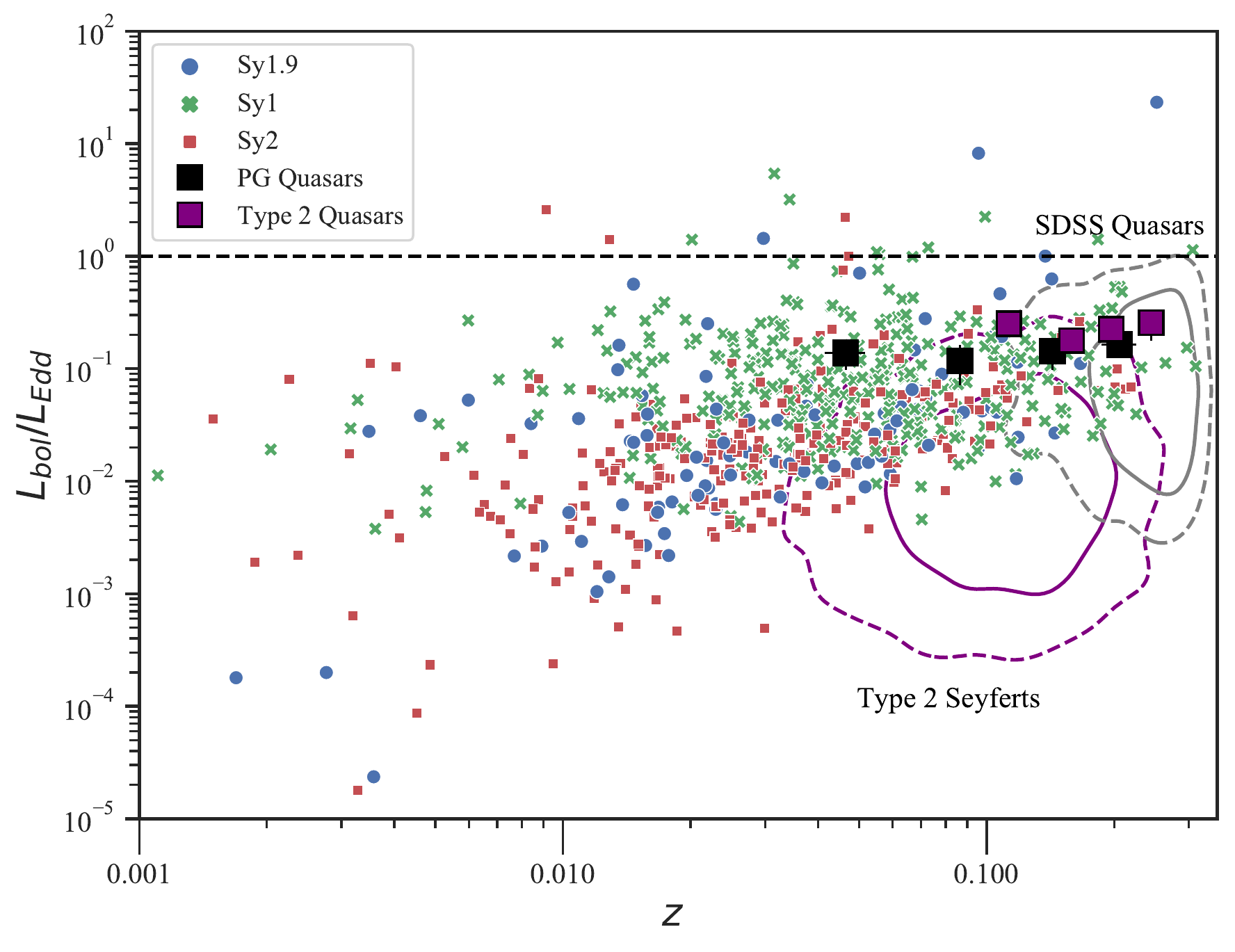}
\includegraphics[width=12cm]{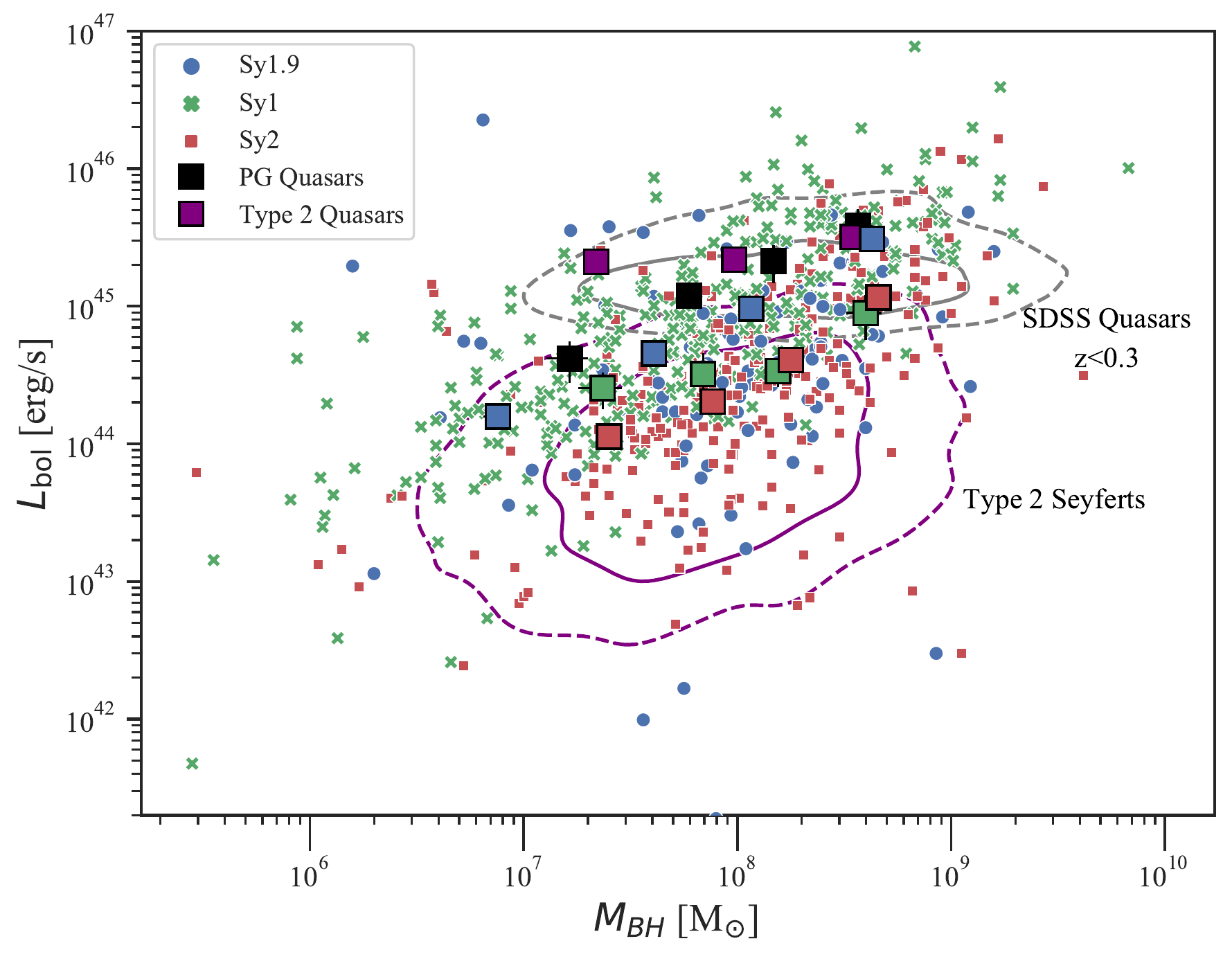}
\caption{The distribution of BASS DR2 AGNs in the Eddington ratio vs. redshift (top) and bolometric luminosity vs. BH mass (bottom) planes.  
The large squares indicate the binned medians for each AGN subclass with redshift (top) and \mbh\ (bottom).  Error bars on the plotted median values are equivalent to 1$\sigma$, calculated based on a bootstrap procedure with 100 realizations. The number of bins was fixed to four constructed to have equal numbers of sources in each bin.  For comparison, we plot SDSS quasars at $z{<}0.3$ \citep[grey contours;][]{Shen:2011:45} and lower-luminosity SDSS Type 2 Seyferts \citep[purple contours;][]{Greene:2005:721}.  The solid and dashed contour covers 68\% and 95\% of the data, respectively. We also plot the median for PG Quasars \citep{Boroson:1992:109} (black squares) and SDSS Type 2 Quasars \citep{Kong:2018:116} selected based on their \OIII\ emission (purple squares).  The BASS AGN have roughly similar BH masses and bolometric luminosities to the different SDSS samples but also extend to lower redshifts and BH masses. We note that there is essentially no overlap between BASS DR2 and these SDSS-based samples of powerful AGNs (see text for discussion).}
\label{fig:lbolmbh}
\end{figure*}

We show the Eddington ratio vs. redshift in the top panel of \autoref{fig:lbolmbh}.  Despite the aforementioned possible biases, Sy 1 sources tend to have higher Eddington ratios (on average) than Sy 1.9 or Sy 2 sources, even when matched in redshift, though this difference decreases towards the highest redshifts in the sample ($z{>}0.05$).  The Sy 1.9 and Sy 2 classes follow the same distribution in \lledd\, rising sharply with redshift from $\lledd\approx10^{-2}$ at $z{=}0.01$ to $\lledd\approx6{\times}10^{-2}$.  
Conversely, the median Eddington ratio of Sy 1s is nearly flat with redshift ($\lledd\approx (7-9){\times}10^{-2}$).

\subsection{Comparison to Other Surveys}
\label{sec:other_surveys}

A comparison to other distant AGN X-ray surveys is shown in \autoref{fig:Lum_xray}.  Due to its all-sky coverage, the BAT AGN include the highest number of sources at ${z}<$0.1, the most luminous unbeamed QSOs at $0.1<z<0.6$, and a population of high-redshift beamed AGN between $z{=}1-3.5$.  In luminosity space the survey overlaps well with medium-deep surveys (e.g. XMM XXL, Stripe 82, and COSMOS) at $0.5<z<3$.

Among all-sky X-ray surveys, it is useful to compare the BAT AGN to the past ROSAT survey in the soft X-ray band \citep[$0.1-2.4$ keV;][]{1982AdSpR...2d.241T}.  The more recent reprocessing of the second ROSAT all-sky survey (2RXS) source catalog \citep{2016A&A...588A.103B} had a flux limit of $\sim10^{-13}$\,\ergcms at 0.1-2.4 keV, with $\sim$130,000 sources. Assuming $\Gamma=1.8$, this flux limit corresponds to $\sim2\times10^{-13}$ over the BAT 14-195\,keV band, which is $\sim$50$\times$ deeper than the 70-month BAT survey.  Of these, 7005 ROSAT sources were cross-matched with the SDSS Data Release 5 \citep{Anderson:2007:313}.  
Due to the soft X-ray sensitivity of ROSAT, the vast majority of these sources (6224/7005, or 89\%) were broad-line AGN whereas the fraction within BASS for such (unbeamed) sources is 42\%. In addition, the median redshift of the (subset of) ROSAT sources was $z=0.42$, which is more than a factor of 10 more distant than the  unbeamed BASS AGN ($z\simeq0.037$).  The BASS overlap with ROSAT is 95\% for Sy\,1 sources, down to 53\% for Sy2, and only 30\% for LINERS \citep[see, e.g.,][for further details]{Oh_DR2_NLR}. A more comprehensive comparison between the 2RXS and the BAT AGNs is also available in \citet{Oh:2018:4}.

The concurrent eROSITA mission and its all-sky survey \citep[0.2-8\,keV][]{2021A&A...647A...1P} are expected to eventually yield a few million AGN and be roughly a factor of 100 times deeper (${\sim}10^{-15}$\,\ergcms) than ROSAT, which means a particularly larger number of higher-redshift sources.  The eROSITA Final Equatorial Depth Survey (eFEDS), with a depth of ${\sim}10^{-14}$\,\ergcms at 0.5-2 keV \citep{2021A&A...647A...5W} over 140\,deg$^2$ provides some early insight into what one could expect for the AGN population to be surveys by eROSITA.  Specifically, 90\% are unobscured (\logNH$<$21.5) and the redshift distribution peaks at around $z\simeq1$ \citep{2021arXiv210614522L}, though redshift determination for the majority of sources is problematic until larger spectroscopic surveys are completed (e.g., via SDSS-V, \citealt{Kollmeier:2017} and/or VISTA/4MOST, \citealt{2021arXiv210614520S})
Thus, we expect that BASS will provide a bright complement of well-understood luminous nearby sources ($z\simeq0.037$) that is less biased with regard to obscuration, but also missing the numerous distant AGN to be detected by eROSITA.


To put the BASS sample in perspective, we also compare it to several other optical surveys of nearby luminous AGN, including the SDSS quasars \citep{Shen:2011:45}, SDSS Sy2 AGN \citep{Greene:2005:721}, the PG quasars \citep{Boroson:1992:109}, and Type 2 Quasars selected using \OIII\ emission \citep{Kong:2018:116}.  Compared to these samples, the BAT-selected AGN are typically found at lower redshifts ($z{<}0.1$).  The Eddington ratios of broad-line (Sy 1) AGN are above the SDSS Sy 2 AGN, consistent with those of SDSS and PG quasars, but below the SDSS Type 2 quasars.  Among Sy 1.9 and Sy 2 types, only the highest redshift BAT sources ($z{>}0.08$) have similar Eddington ratios as the SDSS quasars and PG quasars.  However, the SDSS Type 2 quasars tend to have significantly higher Eddington ratios than BAT Sy1.9 or Sy 2 types, despite overlap in redshift at $z{=}0.1$.  Finally, BASS Sy 2 AGN tend to have higher Eddington ratios than SDSS Sy 2 AGN.

The most luminous quasars in our sample are generally not found in other quasar samples.  We investigated whether the most luminous BAT AGN were selected by the SDSS quasars and Type 2 quasars samples and found virtually no BAT AGN in these samples.  We focused specifically on the range of $z{<}0.3$ and $\Lbol{>}10^{46}\,\ergs$, which includes 18 unbeamed BAT AGN, of which 8 are found within the SDSS footprint.  Of these, only one source (SWIFTJ1547.5+2050 aka 3C323.1) is selected by the SDSS quasar sample.  The other seven AGN were not targeted for SDSS spectroscopy. In five cases the quasar is classified as a star in terms of colors, and in two cases the obscured (Sy2) AGN are classified as galaxies, but no spectra were taken.  

On the other hand, there are 14 SDSS quasars with  \Lbol${>}10^{46}\,\ergs$, which are not detected by the BAT survey. The BAT detection limit at $z{=}0.3$ for 90\% sky coverage is equivalent to \Lbol$=1.1{\times}10^{46}$\,\ergs, assuming a simple conversion of \Lbol$=8{\times}$\Lbat\,\ergps \citep[e.g.,][]{Koss:2017:74}. Hence, all the 14 SDSS quasars should be detected.  It is possible that these undetected quasars may be part of a class of X-ray-weak quasars that have been found by several campaigns \citep[e.g.,][]{1997ApJ...477...93L,Pu_2020}. Alternatively, the single, constant X-ray bolometric correction, rather than a luminosity-dependent one \citep[e.g.,][]{Duras:2020:A73},  may be too low for our sample.Techniques can be used to study known sources at $\sim3\times$ fainter limits within BAT and would be ideal for this population \citep[e.g.,][]{Koss:2013:L26}.   We reserve further discussion for future detailed studies.    
For the SDSS Type 2 Quasars, there is no overlap between the samples, with 26 AGN in the Type 2 Quasars having \Lbol${>}10^{46}\,\ergs$, and 0/26 detected by BAT, when all should be detected.


We further compare the bolometric luminosities and BH masses of the BASS DR2 sample.   Only the most luminous quartile of Sy 1.9 and Sy 2 AGN reaches the average luminosities of the SDSS quasars at similar redshifts ($z{<}0.3$). By comparison, roughly half of the BASS Sy 1 sample occupies similar distributions in bolometric luminosity and BH mass as the SDSS quasars.  The Sy 1 AGN have similar luminosities as the PG quasars, other than the least luminous quartile, but somewhat larger BH masses.  The SDSS Type 2 quasars occupy a region significantly above the BAT Sy 1.9 and Sy 2 AGN in \Lbol,  reached only by the most luminous quartile of the Sy 1 AGN.   

Finally, the BASS sample has a significant number of low-mass BHs ($\mbh{<}10^{7}\,\Msun$) that are not present in any of the other comparison samples. This feature of BASS is due to the higher spectral resolution ($R>2500$) in the optical spectroscopy, which allows us to resolve spectral features tracing smaller BHs; the purely AGN-dominated selection in the $>$14 keV band, which allows to study low-\mbh\ AGNs whose optical emission is host dominated; and the ability to study even the nearest AGN (i.e., $z<0.01$) which are saturated in SDSS imaging and thus excluded from the spectroscopic follow-up.  A small number of nearby ($z{<}0.01$) SMBHs have Eddington ratios as low as $\simeq 10^{-5}$ \citep[e.g., M81,][]{2019MNRAS.488.1199D} and offer an opportunity to study the emission properties of advection-dominated accretion flows (ADAF, \citealt{YuanNarayan14}). These radiatively inefficient accretion flows result in broadband SEDs that are markedly different from those characterizing standard, thin-disk accretion (see, e.g., \citealt{Ryan:2017:199}).

\begin{figure*} 
\centering
\includegraphics[width=12cm]{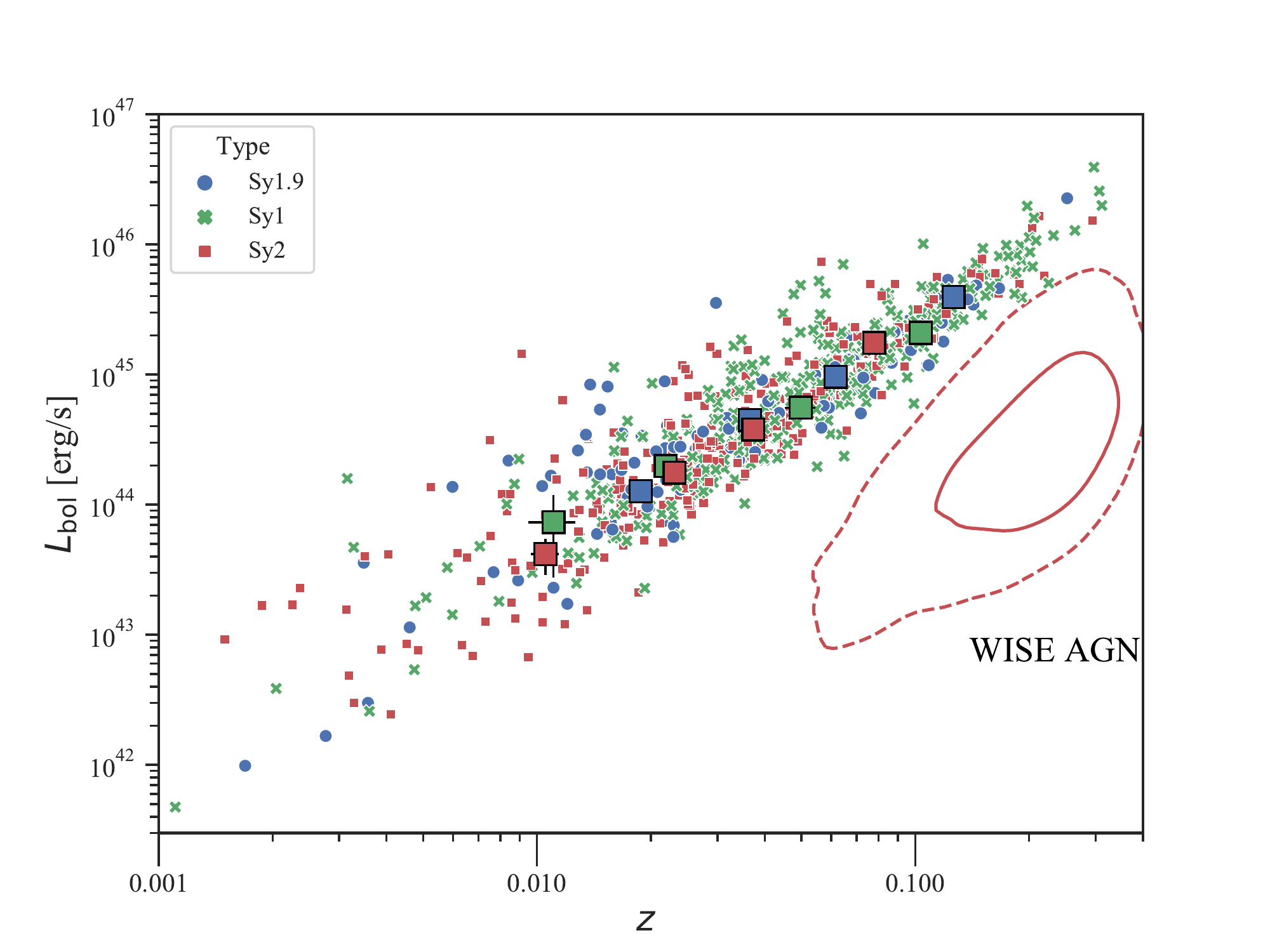}
\includegraphics[width=12cm]{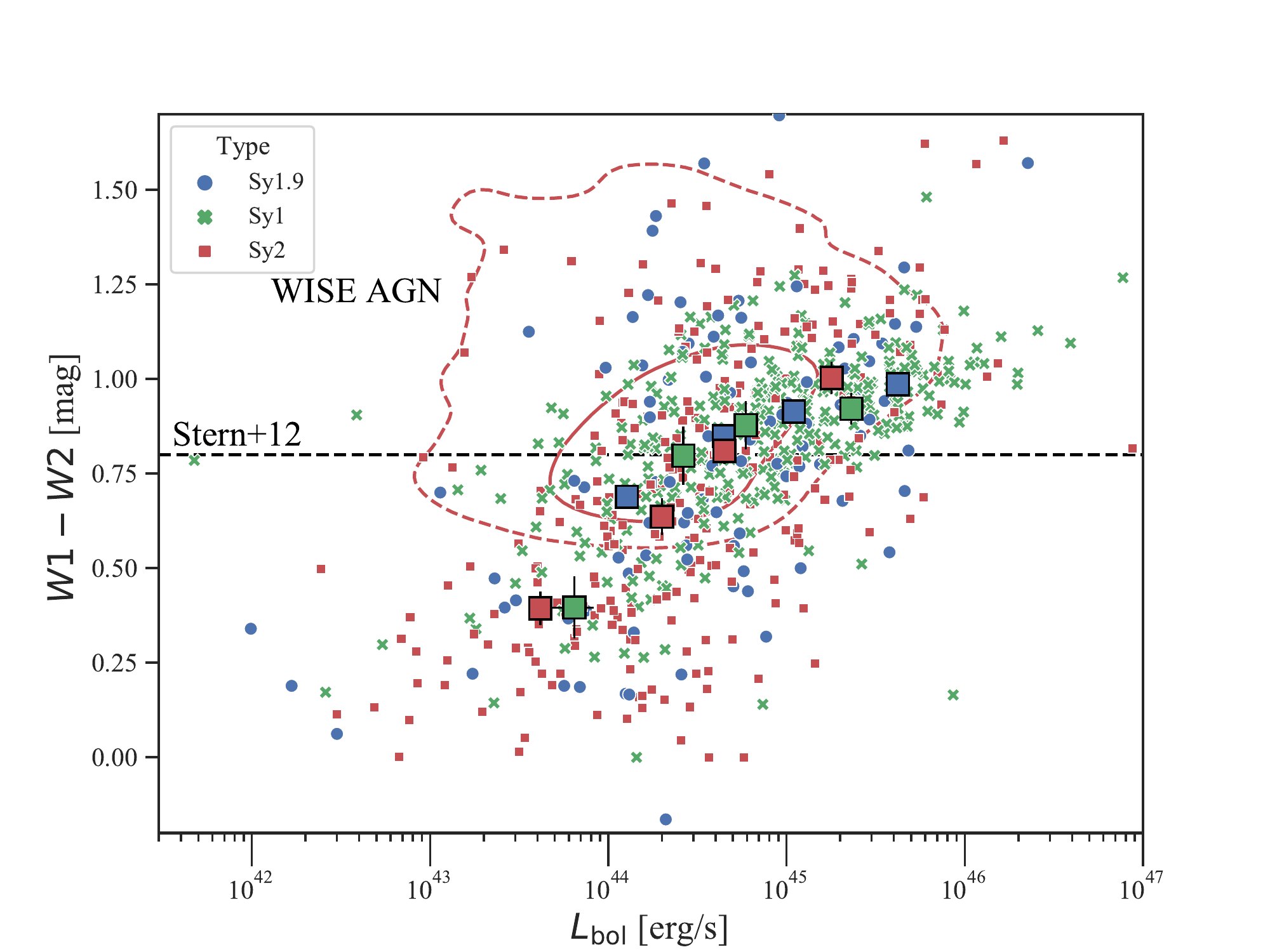}
\caption{The distribution of BASS DR2 in the bolometric luminosity vs. and redshift (top) and WISE colors vs. bolometric luminosity (bottom) planes.  For comparison, we also plot the distributions of WISE-selected AGN at $z{<}0.3$ \cite[red contours]{Assef:2018:23}, with SED fitting and redshifts measured from the SDSS \citep{Barrows:2021:arXiv:2107.02815}. The solid and dashed contours cover 68\% and 95\% of the WISE AGN data, respectively.  A black dashed line indicates the WISE color cut to identify AGN \cite[$W1-W2>0.8$][]{Stern:2012:30}.  The BAT AGN tend to probe similar luminosities as the higher redshift WISE AGN, but at lower redshifts.}
\label{fig:wise}
\end{figure*}

We finally compare the BASS DR2 sample to MIR-selected AGN.  The standard WISE color cut \cite[$W1-W2>0.8$][]{Stern:2012:30} identifies only 56\% of the BAT AGN sample (482/\NAGN).  The fraction of detections is highly dependent on AGN luminosity, with a much higher fraction of luminous AGN detected \citep[\autoref{fig:wise}; see also][]{Ichikawa:2017:74}. 

We also compare the number of BASS AGN in the WISE-selected AGN drawn from the 30,093 deg$^2$ of extragalactic sky in the AllWISE Data Release \citep[][]{Assef:2018:23}.     The AllWISE AGN study by \cite{Assef:2018:23} provides an AGN catalog with 90\% reliability (the ``R90'' catalog), selected purely using the WISE $W1$ and $W2$ bands, but with lower completeness.  However, many of the BAT AGN are excluded by default because they reside in galaxies that are extended in 2MASS, a criterion adopted by the WISE AGN catalog to avoid contamination by separate, resolved parts of nearby galaxies. We find that only about 74/\NAGN\ BASS DR2 sources overlap with the WISE R90 AGN catalog (9\%), including 37 broad line AGN (Sy 1--1.8), 6 narrow line AGN (Sy 1.9--2), 28 beamed broad line AGN, and 3 continuum-dominated blazars (BZB).  This corresponds to a WISE AGN detection fraction of 11.7\% (37/314) for broad line AGN and only 1.5\% for obscured BAT AGN (6/393), though the detection fraction would be higher if extended galaxies were included in the WISE AGN catalog. We note again that the WISE detection fraction is strongly dependent on AGN luminosity, with no BAT AGN with $\Lbol{<}5 \times 10^{44}\,\ergs$ found in the WISE AGN catalog.

When comparing the $\Lbol-z$ distribution of BAT- and WISE-selected AGN, in \autoref{fig:wise}, it appears that the WISE AGN tend toward high redshifts at similar AGN luminosities, due to the requirement that they are point-like in 2MASS.  If we further restrict the WISE AGN to \Lbol${>}10^{46}\,\ergs$ and $z{<}0.3$, there are six WISE AGN above this limit from \citep{Barrows:2021:arXiv:2107.02815}, five of which overlap with the BAT sample. On the other hand, we find 6/18 (33\%) of the unbeamed BAT quasars with $z{<}0.3$ and $\Lbol{>}10^{46}\,\ergs$ are in the WISE all-sky AGN catalog.

In summary, BAT is finding a broad range of nearby AGN in terms of bolometric luminosity, BH mass, Eddington ratio, and particularly obscuration, including a significant population of low \lledd, low \mbh\ sources, with a well-characterized selection function and $>$99\% complete spectroscopic coverage making it a unique legacy sample for future AGN studies.
Given the unique selection criteria provided by BAT and the complete, adaptive optical spectroscopy, it complements other legacy samples of nearby AGN. 

\section{Summary}
\label{sec:summary}

We present here an overview of the BASS DR2, with \Nspec\ optical spectra,  of which \Nspecnew\ are released for the first time, for the \NAGN\ ultra-hard-X-ray-selected AGN in the Swift BAT 70-month sample.  In this special issue, we provide several immediate top-level scientific results and catalogs, including the following:

\begin{enumerate}
    \item A largely statistically complete sample with 99.6\% and 98\% of the brightest \NAGN\ ultrahard X-ray (14-195 keV) selected AGN outside the Galactic plane having measured spectroscopic redshifts and BH mass estimates (respectively). The BH masses are derived from broad emission line \citep{Mejia_Broadlines} or from stellar velocity dispersion measurements \citep{Koss_DR2_sigs}.  
    
    \item The \NAGN\ AGN represent a uniquely complete census of nearby AGN ($z<0.3$), spanning 5-7 orders of magnitude in AGN bolometric luminosity ($\Lbol\sim10^{40}-10^{47}\,\ergs$),  BH mass ($\mbh\sim10^{5}-10^{10}\,\Msun$), Eddington ratio ($\lledd\sim10^{-5}-10$), and obscuration ($\NH\sim 10^{20}-10^{25}\,\cmii$).  These AGN are largely distinct from those found by other surveys, specifically with very little overlap even among nearby SDSS quasars or WISE AGN.
    
    \item A large catalog of emission-line measurements from 3200-10000\,\AA\ \citep{Oh_DR2_NLR} for the \NAGN\ AGN and an additional 233 NIR spectroscopic measurements  \cite[$1-2$\,\micron;][]{denBrok_DR2_NIR,Ricci_DR2_NIR_Mbh}.
    
    \item  The first directly constrained BHMF and ERDF using both unobscured and obscured AGN, in addition to a highly robust determination of the XLF \citep{Ananna_DR2_XLF_BHMF_ERDF}.
    
    \item The significant bias toward underestimation of BH mass when using \hbeta\ or \halpha\ emission in obscured systems  \cite[$\logNH{>}21$; see][]{Mejia_Broadlines,Ricci_DR2_NIR_Mbh}.
    
    \item The ability of MIR emission to recover the X-ray column density \citep{Pfeifle:2022:3}.
    
\end{enumerate}


We hope that these initial results are only the beginning of the legacy value of the BASS project for understanding BH growth in nearby AGN, and that the data products will be of lasting and general usefulness to the broader astronomical community.  There are a variety of studies that can be done using this dataset, such as focusing on SFRs, stellar masses, stellar population ages, dust reddening, metallicities, AGN-driven outflows, weak/faint emission lines, and links to morphological studies --- all of which are not a significant part of the present data release. The broad wavelength coverage of the BASS sample is highly conducive to modeling SED with recent modeling tools \citep[e.g., X-CIGALE; ][]{Yang:2020:740}.  The future DR3 will focus in particular on fainter AGN from the 105-month BAT catalog \citep[][]{Oh:2018:4}, which reaches flux limits 23\% deeper than the 70-month catalog used for DR2, and for which follow-up observations are currently ongoing.   
We encourage the community to engage with the BASS data and team, to maximize the science output of this unique sample and dataset.

\begin{acknowledgments}

We thank the reviewer for the constructive comments that helped us improve the quality of this paper.

BASS/DR2 was made possible through the coordinated efforts of a large team of astronomers, supported by various funding institutions, and using a variety of facilities.

We acknowledge support from NASA through ADAP award NNH16CT03C (M.K.); 
the Israel Science Foundation through grant number 1849/19 (B.T.); 
the European Research Council (ERC) under the European Union's Horizon 2020 research and innovation program, through grant agreement number 950533 (B.T.);
FONDECYT Regular 1190818 (E.T., F.E.B.) and 1200495 (E.T., F.E.B);
ANID grants CATA-Basal AFB-170002 (E.T., F.E.B.), ACE210002 (E.T., F.E.B.) and FB210003 (C.R., E.T., F.E.B.); 
ANID Anillo ACT172033 and Millennium Nucleus NCN19\_058 (E.T.); and Millennium Science Initiative Program  – ICN12\_009 (F.E.B.); an ESO fellowship (M.H.,J.M.); Fondecyt Iniciacion grant 11190831 (C.R.); the National Research Foundation of Korea grant NRF-2020R1C1C1005462 and the Japan Society for the Promotion of Science ID: 17321 (K.O.); Comunidad de Madrid through the Atracción de Talento Investigador Grant 2018-T1/TIC-11035 (I.L.); YCAA Prize Postdoctoral Fellowship (M.B.); from a Clay Fellowship administered by the Smithsonian Astrophysical Observatory and by the BH Initiative at Harvard University, which is funded by grants from the John Templeton Foundation and the Gordon and Betty Moore Foundation (F.P.). 
This work was performed in part at the Aspen Center for Physics, which is supported by National Science Foundation grant PHY-1607611.  
We acknowledge the work done by the 50+ BASS scientists and Swift BAT team to make this project possible.

We acknowledge the various telescopes used in this paper.  
We are tremendously thankful to all the observing and support staff in all the observatories, and their headquarters, for their great assistance in planning and conducting the observations that made BASS/DR2 possible.\\ 
Specifically, BASS/DR2 is based on data obtained through
the European Organisation for Astronomical Research in the Southern Hemisphere (ESO);
the Palomar Observatory;
the Southern Astrophysical Research (SOAR);
the Kitt Peak National Observatory, managed by the US National Science Foundation’s NOIRLab;
the 6.5 m Magellan and the 2.5 m du Pont telescopes located at Las Campanas Observatory, Chile;
the W.M. Keck Observatory at Maunakea, Hawaii;
and the various stages of the Sloan Digital Sky Survey (SDSS).

A full list of observing facilities, program numbers, and their supporting bodies is provided in the main DR2 Catalog paper \cite[][see the acknowledgments there]{Koss_DR2_catalog}.

A significant part of the BASS observations and work took place during the COVID-19 crisis. We thank the healthcare experts in communities around the world, for their tireless efforts to keep us all as safe and healthy as possible.
\end{acknowledgments}

\vspace{5mm}
\facilities{Du Pont (Boller \& Chivens spectrograph), Keck:I (LRIS), Magellan:Clay, Hale (DBSP), NuSTAR, Swift (XRT and BAT), VLT:Kueyen (X-Shooter), VLT:Antu (FORS2), SOAR (Goodman)}

\software{Astropy \citep{Collaboration:2013:A33}, ESO Reflex \citep{Freudling:2013:A96},
           IRAF \citep{Observatories:1999:ascl:9911.002}, Matplotlib \citep{Hunter:2007:90}, 
          Numpy \citep{vanderWalt:2011:22}, Pandas (\url{https://doi.org/10.5281/zenodo.3630805})}

\bibliography{bibfinal,bib_dr2papers,bib_add_here}{}
\bibliographystyle{aasjournal}

\end{document}